%% file: paper.tex
\documentclass[10pt,conference]{IEEEtran}

\usepackage{amsmath} %align and multline environments
\usepackage{amssymb}
\usepackage[ligature, inference]{semantic} %math ligatures and the inference environment
\usepackage{xspace} %\xspace
\usepackage{listings}
\usepackage{graphicx}
\usepackage{url}
\usepackage{multirow}
\usepackage{paralist}

\newcommand{\technique}{path-optimal symbolic execution\xspace}
\newcommand{\techniqueAcronym}{{\sc pose}\xspace}
\newcommand{\TechniqueAcronym}{{\sc Pose}\xspace}
\newcommand{\rep}[1]{\ensuremath{\overline{#1}}}
\NewDocumentCommand{\code}{m}{\text{\texttt{#1}}}

\begin{document}

\title{Path-Optimal Symbolic Execution of Heap-Manipulating Programs}

\author{\IEEEauthorblockN{Pietro Braione}
\IEEEauthorblockA{\textit{University of Milano-Bicocca}\\
Milano, Italy \\
pietro.braione@unimib.it}
\and
\IEEEauthorblockN{Giovanni Denaro}
\IEEEauthorblockA{\textit{University of Milano-Bicocca}\\
Milano, Italy \\
giovanni.denaro@unimib.it}
\and
\IEEEauthorblockN{Luca Guglielmo}
\IEEEauthorblockA{\textit{University of Milano-Bicocca}\\
Milano, Italy \\
luca.guglielmo@unimib.it}
}

\maketitle

\begin{abstract}
Symbolic execution is at the core of many techniques for program analysis and test generation.
Traditional symbolic execution of programs with numeric inputs enjoys the property of forking as many analysis traces as the number of analyzed program paths, a property that in this paper we refer to as path optimality. On the contrary,
current approaches for symbolic execution of heap-manipulating programs fail to satisfy this property, thereby incurring crucial
path explosion effects.
This paper introduces \techniqueAcronym, \technique, a symbolic execution algorithm that originally achieves path optimality against heap-manipulating programs.
We formalize the \techniqueAcronym algorithm 
and experiment it 
against a benchmark of  programs  that take data structures as inputs, supporting 
the potential of  \techniqueAcronym for improving on the state of the art of symbolic execution of heap-manipulating programs.
\end{abstract}

\mathlig{->}{\rightarrow}
\mathlig{~>}{\rightsquigarrow}
\mathlig{=>}{\Rightarrow}
\mathligson

\input{intro}
\input{motivating}

\input{language}

\input{experiments}
\input{related}

\input{conclusions}

\bibliographystyle{splncs04}
\bibliography{bibliography}

\appendix
\input{formal_long}

\end{document}

%% file: intro.tex
\section{Introduction} \label{sec:intro}

Symbolic execution is a well known and reputed program analysis technique. Since the seminal work of King and Clarke in the seventies~\cite{king_symbolic_1976,clarke_program_1976}, it has been increasingly gaining momentum in both academia and industry~\cite{Pasareanu:survey:JSTTT2009,cadar_symbolic_2013,braione_software_2014,SurveySymExec-CSUR18}, with a plethora of effective applications for program verification~\cite{braione:enhancing:esecfse:2013,sinha:symbolic:fmcad:2008,laziersharp:sefm:2007,coen:symbolic:fse:2001,deng_bogor_kiasan_2006}, test case generation~\cite{godefroid:dart:pldi:2005,tillmann_pex_2008,li_klover_2011,vivanti_search-based_2013,braione_symbolic_2015,bucur:parallel:eurosys:2011,anand_jpf-se_2007,pasareanu:mixed:issta:2011,staats:parallel:issta:2010}, worst case time estimation~\cite{burnim_wise_2009,aquino_worst-case_2018,luckow_symbolic_2017}, fuzzing~\cite{stephens2016driller,symbex:fuzzing:2011}, vulnerability analysis~\cite{he:se:smartcontract:2020,zheng2022park}, and many others~\cite{filieri:probabilistic:ase:2015,yang:incremental:tosem:2014,yang2012memoized}.

In a nutshell, symbolic execution consists in   simulating the semantics of programs against symbolic inputs, representative of all possible input values, thereby synthesising the conditions on the input values that make given program
paths execute. It exploits these \emph{path conditions} with constraint solvers (e.g., Z3~\cite{de_moura_z3_2008}) to reason on the feasibility of the program paths~\cite{cadar_klee_2008,baluda2016bidirectional,burnim_heuristics_2008,braione_jbse_2016,anand_jpf-se_2007}.

To analyze \emph{heap-manipulating programs}, which take inputs that can reference data structures in the heap memory, symbolic execution  shall cope with both numeric and reference-typed inputs.
Since the seminal work of Visser et al.~in 2003~\cite{khurshid:tacas:2003}, the classic approach is to augment traditional symbolic execution with \emph{lazy initialization}~\cite{visser:test:issta:2004,anand_jpf-se_2007,braione_jbse_2016,geldenhuys2013bounded,rosner2015bliss,braione_symbolic_2015}. Upon accessing a  reference-typed input (say parameter \code{r}) for the first time, lazy initialization systematically enumerates the assumptions on how \code{r} can relate to the data structures in the initial heap:  \code{r} could be \code{null}, or alias of any other type-compatible input reference,
or point to a distinct input data structure possibly containing further symbolic inputs. For each assumption, lazy initialization forks a separate analysis trace, 
to analyze how the assumption impacts at  the next steps of the program. This embodiment of symbolic execution 
is referred to as \emph{generalized symbolic execution}~\cite{khurshid:tacas:2003,siddiqui:staged:sac:2012}. 

Lazy initialization is elegant and simple, but it makes symbolic execution give up
the property of traditional symbolic execution of forking exactly an analysis trace for each analyzed program path. In this paper we refer to this property as \emph{path optimality}. It requires that \emph{the analysis traces forked during symbolic execution are isomorphic to the analyzed program paths.}
Conversely, with lazy initialization, the symbolic execution of each program path forks a number of analysis traces combinatorial in the number of assumptions made while accessing  heap-related inputs. This generally exacerbates, possibly  to  large extents (see Section~\ref{sec:experiments}), the path explosion issues incurred by symbolic execution.  At the state of the art, while researchers investigated some ways  of  improving the efficiency of the lazy initialization algorithm~\cite{deng_bogor_kiasan_2006,laziersharp:sefm:2007,kiasan:jase:2012} (as we survey in  Section~\ref{sec:motivating}), no existing variant of generalized symbolic execution achieves path optimality. 

In this paper we introduce a novel \emph{\technique} (\techniqueAcronym) algorithm for heap-manipulating programs. When the program accesses the heap via an input reference, \techniqueAcronym models the result with a symbolic formula that embeds all relevant assumptions on aliasing and distinctness between the input references as a symbolic expression, by using \emph{if-then-else} expressions. Thus, it represents the relevant assumptions precisely, but avoids forking the analysis at any step but actual decision points of the program. The original contributions of this paper are: 
\begin{enumerate}
    \item A formalization of the \techniqueAcronym algorithm for a tiny but realistic object-oriented language;
    \item A prototype implementation of the \techniqueAcronym algorithm into a symbolic executor for programs in Java bytecode;
    \item An experimental comparison of \techniqueAcronym against lazy initialization over a benchmark of subject programs from the SBST Java  testing tool competition~\cite{panichella:comp:sbst:2017}.
\end{enumerate}
 Our experiments confirm that \techniqueAcronym achieves path optimality,  reducing the number of analysis traces of several orders of magnitude, with respect to classic lazy initialization. Furthermore, thanks to path optimality,  \techniqueAcronym significantly reduces the time spent for constraint solving, and enables major performance improvements in test generators based on symbolic execution.  

This paper is organized as follows. Section~\ref{sec:motivating} exemplifies classic lazy initialization on a set of sample programs, to motivate our work and make the paper self-contained. Section~\ref{sec:language}  introduces \techniqueAcronym by example, and presents the \techniqueAcronym algorithm. Section~\ref{sec:experiments} discusses the experiments that we executed with our prototype tool, to evaluate \techniqueAcronym. Section~\ref{sec:related} discusses related work. Section~\ref{sec:conclusions} outlines the conclusions of the paper and our plans for further research on \techniqueAcronym.

%% file: motivating.tex
\section{Motivating Examples} \label{sec:motivating}
In this section we discuss the suboptimality of the state-of-the-art symbolic execution algorithms in coping with heap-manipulating programs.
The mainstream approaches by which symbolic execution is  extended to heap-manipulating programs are \emph{lazy initialization}~\cite{khurshid:tacas:2003} and \emph{lazier\# intialization}~\cite{kiasan:jase:2012}. Below, we introduce these two classical algorithms, and discuss their limitations (that in turn motivate our work) with reference to the three sample programs in Figure~\ref{fig:sample:programs}.

\begin{figure*}[t]
    \centering \scriptsize

\addtolength{\tabcolsep}{10pt}
\begin{tabular}{ccc}

\begin{lstlisting}[language=java,numbers=left,tabsize=2,
    stepnumber=1, basicstyle=\scriptsize]
class Sample {
 Object data;
 void swap(Sample s) {    
  if (s != null) {
   Object d = this.data;
   this.data = s.data;
   s.data = d;
  }
 }
}
\end{lstlisting} &
\begin{lstlisting}[language=java,numbers=left,
    stepnumber=1, basicstyle=\scriptsize]
class Sample {
 int val;
 Sample s0,s1,s2;
 int sum() {  
  int sum = val;
  sum += s0.val;
  sum += s1.val;
  sum += s2.val;
  return sum;
 }
}
\end{lstlisting} &

\begin{lstlisting}[language=java,numbers=left,tabsize=2,
    stepnumber=1, basicstyle=\scriptsize]
class Sample {
 int max = 4;
 Sample next;
 boolean hasNull() {
  Sample s = next;
  int i = 1;
  while (s != null
   && i <= max) {
    s = s.next;
    i = i + 1;
  }
  return s == null;    
 }
}
\end{lstlisting}\\
(a) & (b) & (c)\\
\end{tabular}
    \caption{Sample programs}
    \label{fig:sample:programs}
\vspace{-10pt}
\end{figure*}

\subsection*{Example 1: Swap Program}
In the program of Figure~\ref{fig:sample:programs}~(a), method \texttt{swap} of class \texttt{Sample} swaps the data within the receiver object (accessed in field \texttt{data}) with the data \texttt{s.data} within another \texttt{Sample} object that the method receives as parameter. The code of method \texttt{swap} consists of two possible program paths: if parameter \texttt{s} is \texttt{null},  the program  skips the swap behavior; otherwise, the swapping indeed takes place. According to path optimality, we would like symbolic execution to accomplish the symbolic analysis of this program 
as a symbolic computation tree with exactly 2 traces,  with path conditions \code{s == null} and \code{s != null}, respectively. 

\paragraph{Lazy initialization} Lazy initialization works as follows: When a symbolic reference originating from an input is used for the first time during symbolic execution, a case analysis is performed and the current symbolic state is refined (``split'') based on the result of this analysis. The possible cases are:
\begin{inparaenum}[(i)]
\item the input reference may be null, or 
\item it may point to an object of the initial heap that is being accessed for the first time during symbolic execution (often referred to as a \emph{fresh} object), or 
\item it may point to any of the type-compatible objects that were assumed at earlier symbolic execution steps (referred to as  \emph{aliases}).
\end{inparaenum}

At line 4 of program \texttt{swap} in Figure~\ref{fig:sample:programs}~(a), where the symbolic reference  \code{s} is first used to access \code{s.data}, we shall consider three possible cases: \code{s == null}, \code{s} aliases the receiver object \code{this} (the \code{this} reference is initialized before all the others), or \code{s} points  to a fresh initial object that differs from \code{this}. In similar fashion, upon analyzing lines 5 and 6, we must separately consider the possible initial values of reference \code{this.data} (i.e., \code{null}, alias of either \code{this} or \code{s}, or a pointer to a fresh object) and reference \code{s.data} (i.e., \code{null}, alias of either \code{this}, \code{this.data} or \code{s}, or a pointer to a fresh object), respectively. 

An actual run of symbolic execution equipped with lazy initialization reveals that the analysis of program \texttt{swap} unfolds the state space of the program as a symbolic computation tree with 21 traces, which is largely suboptimal with respect to our expectation of only 2 traces.

\paragraph{Lazier\# Initialization} Deng et al.~\cite{kiasan:jase:2012} proposed lazier\# initialization as an optimization of the lazy initialization algorithm. The idea is to delay the enumeration of the possible values of an input reference, 
until the program executes  operations  against the reference. 

Initially, the technique  represents input references by means of abstract symbolic values. It uses these abstract values directly, that is, without enumerating the alternative initializations, whenever the program simply assigns the value of a reference to a variable or to a parameter. For instance, at line 5 of program \code{swap} in Figure~\ref{fig:sample:programs}~(a), the symbolic value of \code{this.data} can be assigned to the local variable \code{d}, without enumerating the possible initializations thereby. 

The program statements that simply compare a symbolic reference with null, cause the enumeration of the symbolic reference to either \code{null}, or a \emph{symbolic heap location}, i.e., a symbolic value that represents a (non-\code{null}) location in the heap. This further mitigates path explosion. For instance, at line 4 of program \code{swap}, lazier\# initialization produces only 2 traces. 

Eventually, however, lazier\# initialization may fall back to baseline lazy initialization, to execute program statements that shall access the specific heap location associated with a reference. For instance, at line 6 of program \code{swap}, accessing field \code{s.data} requires symbolic execution to explicitly discriminate whether the non-null value of reference \code{s} is or is not alias of \code{this}.  

In total, for program \code{swap} lazier\# initialization represents the state space of the program as a symbolic computation tree with 3 traces, which is only
slightly suboptimal with respect to our expectation of only 2 traces.

\subsection*{Example 2: Sum Program}
We further discuss the limits of the optimizations brought by lazier\# initialization, by exemplifying the symbolic execution of the \code{sum} program in Figure~\ref{fig:sample:programs}~(b). This is a straight-line program that sums the value field of three nodes. We expect that a path-optimal symbolic execution of this program accomplishes the analysis with exactly 1 trace, as there are no decision points in the programs.

Unfortunately, for this program, both lazy initialization and lazier\# initialization yield exactly the same suboptimal result. In fact, as the program accesses field \code{val} within the data structures associated with the inputs \code{s0}, \code{s1} and \code{s2}, symbolic execution shall determine the exact locations of such fields, which requires to enumerate all possible initializations of those references in turn. To read the value of field \code{s0.val} (line 6), the cases \code{s0 == null}, \code{s0 == this}, and \code{s0 fresh} shall be all considered. Accessing \code{s1.val} and \code{s2.val} requires homologous initializations and additional cases of \code{s1 == s0}, \code{s2 == s0}, and \code{s2 == s1}. A run of symbolic execution equipped with either the lazier\# algorithm or standard lazy initialization suboptimally generates a symbolic computation tree with 23 traces in both cases.

\subsection*{Example 3: HasNull Program}
The suboptimality of classic generalized execution may exacerbate the path-explosion issues of symbolic execution to large extent. We exemplify this phenomenon with reference to the \code{hasNull} program of Figure~\ref{fig:sample:programs}~(c). It scans a sequence of \code{Sample} nodes in search for a \code{null} terminator within \code{max} nodes, where \code{max} is a constant set to 4 in the example. Since each evaluation of the loop condition at line 7 (\code{s != null}) generates two alternatives, one that exits the loop and returns \code{true}, and one that enters the next iteration, respectively, and since the fifth evaluation of of the loop condition at line 8 (\code{i <= max}) exits the loop in any case, a path-optimal symbolic execution must yield 6 traces.

According to lazy initialization, accessing \code{this.next} at line 5 requires to consider three alternatives, i.e., \code{this.next == null}, \code{this.next == this} and \code{this.next fresh}, respectively. Each access to \code{s.next} at line 9 further requires similar initializations, plus additional alias cases with respect to the incrementally considered fresh objects. Lazier\# initialization achieves limited improvements, only differing in not enumerating the cases other than \code{s.next == null} at the fifth nesting level. In the end, the symbolic computation tree consists of 21 traces and 16 traces with lazy and lazier\# initialization, respectively.

Generalizing the example for other values of the bound \code{max}, 
a path-optimal symbolic execution should analyze $O(max)$ traces, while lazy and lazier\# initialization lead to analyzing $O(max^2)$ traces. Changing line 2 as \code{max=10} yields 78 traces with lazy initialization, instead of 12 path-optimal traces.

%% file: language.tex
\section{Path-Optimal Symbolic Execution} \label{sec:language}

\TechniqueAcronym is a symbolic execution algorithm for heap-manipulating programs that realizes the notion of path optimality introduced in Section~\ref{sec:intro}, 
forking analysis traces only when analysing distinct paths of the program control flow graph.

\TechniqueAcronym grounds on the intuition that most accesses that a program makes to the data structures in the heap, e.g., reading and writing values from fields of input objects, serve in the scope of sequential (non-branching) computations, such as, for computing other values, assigning some variable, dereferencing further objects, and so forth. To accomplish path-optimal symbolic execution, \techniqueAcronym aims to represent the effects of those instructions with comprehensively precise symbolic expressions, but avoiding to spawn distinct path conditions for any instruction that does not correspond to an actual decision point of the program. This is the distinctive characteristic of \techniqueAcronym with respect to the previous techniques like lazy and lazier\# initialization that we surveyed in Section~\ref{sec:motivating}. Indeed, while programs  access the heap to read and write values quite often, only occasionally those values will be participating in decision points that select the execution of distinct program paths. 

In this section, we first provide an informal overview of how \techniqueAcronym handles the representation of the symbolic states when executing program instructions that access the data structures in the initial heap. We organize this part of the presentation around simplified examples, aiming to introduce the core technical means that comprise \techniqueAcronym. Then, we present the \techniqueAcronym algorithm formally. 

\subsection{Overview of \techniqueAcronym}

\begin{figure*}[t]
\scriptsize
\begin{tabular}{p{10pt}cp{5pt}c}
&\begin{lstlisting}[language=java,numbers=left,
    stepnumber=1,basicstyle=\scriptsize]
class A {int f;}
int p1(A a0, A a1, A a2) {
    int v = a0.f + a1.f + a2.f;
    if (a0.f != a1.f)
        abort();
    return v - a1.f - a2.f;
}
\end{lstlisting} 
& &
\begin{tabular}{p{0.01\textwidth}p{0.5\textwidth}} 
\(\texttt{a0}\) & \(:= Y_0 \mapsto (f \mapsto X_0)\)\\

\(\texttt{a1}\) & \(:= Y_1 \mapsto (f \mapsto \code{ite}(Y_1 = Y_0, X_0, X_1))\)\\

\(\texttt{a2}\) & \(:= Y_2 \mapsto (f \mapsto \code{ite}(Y_2 = Y_1, \code{ite}(Y_1 = Y_0, X_0, X_1),\) \\
& \(\code{ite}(Y_2 = Y_0, X_0, X_2)))\)\\

\(\texttt{v}\) & \(:= X_0 + \code{ite}(Y_1 = Y_0, X_0, X_1) + \code{ite}(Y_2 = Y_1, \code{ite}(Y_1 = Y_0, X_0, X_1),\) \\
& \(\code{ite}(Y_2 = Y_0, X_0, X_2))\)\\

\(\Sigma\) & \(:= Y_0 \ne \code{null} \land Y_1 \ne \code{null} \land Y_2 \ne \code{null}\)
\end{tabular}\\
~\\

\scriptsize & (a) The sample program \texttt{p1} & & \scriptsize(b) Symbolic state after  line 3 of \texttt{p1}

\end{tabular}
    \caption{Example on how \techniqueAcronym handles reading values from the heap}
    \label{fig:sampleprog1}
\vspace{-10pt}
\end{figure*}

Let us consider the sample program \texttt{p1} in Figure~\ref{fig:sampleprog1}~(a) that accesses the input heap to read values. At line 3, the program dereferences the three objects referred by the inputs \texttt{a0}, \texttt{a1} and \texttt{a2}. If we exclude (for the moment) that those inputs could be \code{null} references, the computation at line 3 is strictly sequential: it reads the values from the field \texttt{f} of the three input objects, one after the other, and sums them up incrementally. The result of line 3 does indeed depend on the possibility that  \texttt{a0}, \texttt{a1} and \texttt{a2} can be mutually alias of each other, but these possible alias relations do not drive the control to different program branches. 

Aiming to path optimality, in \techniqueAcronym the execution of program \texttt{p1}, line 3,  leads to a single symbolic state that comprehensively represents the possible aliasings, as shown in Figure~\ref{fig:sampleprog1}~(b):
Variable \texttt{a0} refers to a symbolic location $Y_0$, in turn associated with an object of the input heap in which the field \texttt{a0.f} has a symbolic integer value $X_0$. 
Variable \texttt{a1} refers to a symbolic location $Y_1$ associated with an object in which field \texttt{a1.f} has two possible values, thereby defined with an \emph{if-then-else} expression (\emph{ite} expression), i.e., the expression \(\code{ite}(Y_1 = Y_0, X_0, X_1)\): It represents the fact that the value of the field is the symbol $X_0$ if $Y_1$ and $Y_2$ represent the same concrete location, or \texttt{a1.f} has another independent symbolic value, $X_1$, otherwise. Variable \texttt{a2} refers to a symbolic location $Y_2$, and thereby to an object where the value of \texttt{a2.f} is modeled by a composition of ite expressions that represent the possible alias relations between the locations $Y_2$, $Y_1$ and $Y_0$. Variable \texttt{v} evaluates to the sum of the symbolic values of   \texttt{a0.f}, \texttt{a1.f} and \texttt{a2.f}. 

The path condition ($\Sigma$ in Figure~\ref{fig:sampleprog1}~(b)) records only that the three symbolic locations $Y_0$, $Y_1$ and $Y_2$ correspond to actual concrete memory locations (i.e., not to \code{null}) which is in fact the only condition for the program to execute line 3 and proceed on. As a matter of fact, when executing program \texttt{p1}, any concrete assignment of \texttt{a0}, \texttt{a1} and \texttt{a2} with non-\code{null} values will make the program execute through line 3,
regardless how these inputs may alias between each other.

The only branching point of the program \texttt{p1} in Figure~\ref{fig:sampleprog1}~(a) is at line 4. It depends on the condition \texttt{a0.f != a1.f}, which the \techniqueAcronym algorithm will render as in classic symbolic execution, i.e., by evaluating the condition with respect to the symbolic values of these fields. Since the values of \texttt{a0.f} and \texttt{a1.f} are $X_0$ and $\code{ite}(Y_1 = Y_0, X_0, X_1)$, respectively, this leads to line 5 being guarded by the path condition \(Y_0 \ne \code{null} \land Y_1 \ne \code{null} \land Y_2 \ne \code{null} \land X_0 \ne \code{ite}(Y_1 = Y_0, X_0, X_1)\). A constraint solver can straightforwardly compute that this path condition is satisfiable only if $Y_1$ and $Y_0$ are different. 
Similarly the return value at line 6 can be symbolically computed to be $X_0$, with the path condition including the clause \(X_0 = \code{ite}(Y_1 = Y_0, X_0, X_1)\), satisfiable both if $Y_1$ and $Y_0$ are equal and if they are not.
We observe that the path conditions to execute line 5 and line 6 do not depend on possible alias relations of $Y_0$ and $Y_1$ with respect to $Y_2$, which is in fact what we expect for this program. 

We now elaborate on how \techniqueAcronym models the write accesses to the heap. We refer to the sample program \texttt{p2} in Figure~\ref{fig:sampleprog2}~(a) that assigns the values of the fields \texttt{b2.f}, \texttt{b1.f} and \texttt{b0.f} (lines 3--5) thus updating the initial values of those fields. The condition at line 6 can hold true only if the references \texttt{b2}, \texttt{b1} and \texttt{b0} are mutually aliases, since in that case \texttt{b2.f}, \texttt{b1.f} and \texttt{b0.f} all refer to the same field, which is set to 0 after the latest assignment at line 5. In the same spirit of the previous example, \techniqueAcronym  models the assignments at lines 3--5 by taking track of the possible alias conditions, but without spawning multiple path conditions. As above, for now, we consider that the three input references are not null. 

\begin{figure}[t!]
\scriptsize
\begin{tabular}{c}
\begin{tabular}{p{10pt}p{0.53\textwidth}}
&\begin{lstlisting}[language=java,numbers=left,
    stepnumber=1, basicstyle=\scriptsize, aboveskip=0pt, belowskip=0pt, belowcaptionskip=0pt, abovecaptionskip=0pt, captionpos=b]
class B {int f;}
void p2(B b0, B b1, B b2){
    b2.f = 2;
    b1.f = 1;
    b0.f = 0;
    if (b0.f + b1.f + b2.f == 0) {
        abort();
    }
}
\end{lstlisting} \\
& (a) The sample program \texttt{p2}
 \\~\\~\\
& \(\texttt{b2} := Y_2 \mapsto (f \mapsto 2)\)\\

& \(\texttt{b1} := Y_1\)\\

& \(\texttt{b0} := Y_0\)\\

& \(\Sigma := Y_2 \ne \code{null}\)\\~\\

& (b) The symbolic state after  line 3 of \texttt{p2}\\~\\~\\
%\end{tabular}
%\begin{tabular}{p{0.43\textwidth}}
& \(\texttt{b2} := Y_2 \mapsto (f \mapsto \code{ite}(Y_2 = Y_1, 1, 2))\)\\

& \(\texttt{b1} := Y_1 \mapsto (f \mapsto 1)\)\\

& \(\texttt{b0} := Y_0\)\\

& \(\Sigma := Y_2 \ne \code{null} \land Y_1 \ne \code{null}\)\\~\\

& (c) The symbolic state after  line 4 of \texttt{p2} \\~\\~\\

& \(\texttt{b2} := Y_2 \mapsto (f \mapsto \code{ite}(Y_2 = Y_0, 0, \code{ite}(Y_2 = Y_1, 1, 2)))\)\\

& \(\texttt{b1} := Y_1 \mapsto (f \mapsto \code{ite}(Y_1 = Y_0, 0, 1))\)\\

& \(\texttt{b0} := Y_0 \mapsto (f \mapsto 0)\)\\

& \(\Sigma := Y_2 \ne \code{null} \land Y_1 \ne \code{null} \land Y_0 \ne \code{null}\)\\~\\

& (d) The symbolic state after  line 5 of \texttt{p2} \\
\end{tabular}
\end{tabular}
    \caption{Example on how \techniqueAcronym handles assigning values in the heap}
    \label{fig:sampleprog2}
%\vspace{-12pt}
\end{figure}

The symbolic execution of program \texttt{p2}, line 3, leads to the symbolic state shown in Figure~\ref{fig:sampleprog2}~(b): Variable \texttt{b2} refers to the symbolic location $Y_2$, associated with a data structure in which field \texttt{f} is set to 2 as result of the assignment, and both \texttt{b1} and \texttt{b0} are modeled as unaccessed symbolic references, not yet initialized to a corresponding object. The path condition $\Sigma$ records that $Y_2$ must be not null, which is the requirement for accessing the field \texttt{b2.f}. 

Then, when executing line 4 \techniqueAcronym updates the symbolic state as shown in Figure~\ref{fig:sampleprog2}~(c): The value of variable \texttt{b1} reflects the assignment of \texttt{b1.f} to 1, and the value of \texttt{b0.f} reflects that \texttt{b0.f} would be updated, if the symbolic locations $Y_1$ and $Y_2$ were aliases. In \techniqueAcronym, when updating a field in an object associated to a symbolic location (e.g., $Y_1$), we also update  the same field in the type-compatible objects stored at  symbolic locations that belong to the same symbolic state. In the example, since $Y_1$ and $Y_2$ represent type-compatible input references, they might be alias in the input state: if so, the assignment of $Y_1.f$ entails also the assignment of $Y_2.f$. \TechniqueAcronym models $Y_2.f$ as the expression \(\code{ite}(Y_2 = Y_1,1,2)\) that reflects the assignment to 1, if $Y_2$ and $Y_1$ are aliases, or the previous value of the field, 2, if $Y_2$ and $Y_1$ differ.
The path condition records that $Y_1$ is not null, for \texttt{b1.f} to be successfully accessed. 

Proceeding in this way, after line 5 we get the symbolic state shown in Figure~\ref{fig:sampleprog2}~(d): Variable \texttt{b0} refers to the symbolic location $Y_0$ with field \texttt{f} assigned to 0, and the type-compatible input objects pointed by the symbolic locations $Y_1$ and $Y_2$ are updated with ite expressions to model the effects of the possible alias relations between $Y_1$ and $Y_0$, and between $Y_2$ and $Y_0$, respectively. The path condition now records that also $Y_0$ must be not null.

Eventually, at line 6, \techniqueAcronym proceeds by evaluating the condition for the branching point based on the symbolic state computed as above. This leads to computing the path condition for the execution at line 7 based on symbolic values of the three fields, which is then:
\(Y_2 \ne \code{null} \land Y_1 \ne \code{null} \land Y_0 \ne \code{null} \land \code{ite}(Y_2 = Y_0, 0, \code{ite}(Y_2 = Y_1, 1, 2)) + \code{ite}(Y_1 = Y_0, 0, 1) + 0 = 0.\)
This path condition is satisfiable when $Y_0$, $Y_1$ and $Y_2$ are alias of the same non-null concrete location, which is what characterizes the input states that execute up to line~7.

Finally, going back to the possibility that the inputs of programs \texttt{p1} and  \texttt{p2} could be null references, we remark that those cases correspond to distinct control flow paths of the two programs, namely, the paths in which the two programs raise specific exceptions for the given null-pointers.  
Thus, \techniqueAcronym analyzes those paths in separate traces as with classic lazy initialization.

\subsection{Formalization of \techniqueAcronym}
We formalized the \techniqueAcronym algorithm for imperative, object-oriented programming languages. Without loss of generality, our formalization refers to a tiny language that models a program as a sequence of classes and their methods, with single-class inheritance. The tiny language includes instructions for instantiating new objects, reading and assigning their fields, structuring the control flow with if-statements and method calls, and operating with expressions on references, integers and booleans. The tiny language does not have loop statements, but  iterative computations can be anyway defined via recursion. 

\begin{figure}[t!]
\centering\scriptsize
\(
\inference[\scriptsize Getfield~(c1):]{
Y \in \mathrm{dom}(H) \quad f \in \mathrm{dom}(H(Y)) \quad H(Y)(f) \neq \bot
}{
H \ \Sigma \ Y.f -> H \ \Sigma \ H(Y)(f)
}
\)
\\~\\~\\
\(
\inference[\scriptsize Getfield~(c2):]{
H \ \Sigma \ \sigma_1.f => H'_1 \ \Sigma'_1 \ \sigma'_1 \quad H \ \Sigma \ \sigma_2.f => H'_2 \ \Sigma'_2 \ \sigma'_2 \\
H' = mergeHGf(H'_1, H'_2, f, \sigma) \quad \Sigma' = mergePC(\sigma, \Sigma'_1, \Sigma'_2)
}{
H \ \Sigma \ \code{ite(}\sigma, \sigma_1, \sigma_2\code{)}.f -> H' \ \Sigma' \ \code{ite(}\sigma, \sigma'_1, \sigma'_2 \code{)}
}
\)
\\~\\~\\
\(
\inference[\scriptsize Getfield~(r1):]{
Y \notin \mathrm{dom}(H) \quad H' = H, Y \mapsto (fields(class(f)) \mapsto \bot) \\ \Sigma' = \Sigma, \neg Y = \code{null}
}{
H \ \Sigma \ Y.f ~> H' \ \Sigma' \ Y.f}
\)
\\~\\~\\
\(
\inference[\scriptsize Getfield~(r2):]{
Y \in \mathrm{dom}(H) \quad H(Y)(f) = \bot \quad Z\ \text{fresh} \\
H' = H[Y \mapsto H(Y) [f \mapsto assume(H, Y, f, Z)]] \\
\Sigma' = \Sigma, Y.f = Z
}{
H \ \Sigma \ Y.f ~> H' \ \Sigma' \ Y.f
}
\)
\caption{Refinement and computation transition relations for getfield expressions}
\label{fig:semantics:getters}
\vspace{-15pt}
\end{figure}

For any expression $\eta$ of the tiny language, where $\eta$ could be for example a getfield or a putfield expression, we formalize  the small-step operational semantics as 
\[
\inference[Step:]{
H \ \Sigma \ \eta ~>^* H' \ \Sigma' \ \eta \quad H' \ \Sigma' \ \eta -> H'' \ \Sigma'' \ \eta''
}
{
H \ \Sigma \ \eta => H'' \ \Sigma'' \ \eta''
}
\]
where $H, H', H''$ are heaps, $\Sigma, \Sigma', \Sigma''$ are path conditions, and $\eta''$ is the result obtained by reducing the expression $\eta$. The step relation ($=>$) is the suitable composition of two transition relations: the (reflexive-transitive closure of the) \emph{refinement} transition relation $~>$, which augments $H \ \Sigma$ with the relevant assumptions needed to reduce the expression at hand, leading to $H' \  \Sigma'$; and the \emph{computation} transition relation $->$, which further processes the symbolic configuration to reflect the side effects of reducing the expression $\eta$ to $\eta''$, leading to the final configuration
$H''\ \Sigma''\ \eta''$.

Below we discuss the formalization of \techniqueAcronym in particular for the expressions that read and write the content of object fields, i.e., expressions like $\sigma.f$ and $\sigma.f := \sigma'$, where $\sigma$ is either a concrete or a symbolic reference value. As explained above, those are the expressions
for which the symbolic execution according to \techniqueAcronym distinctively differs from generalized symbolic execution based on classic lazy initialization. Furthermore, we also discuss the handling of conditional statements that predicate over references and method calls, whose semantics relates to subexpressions that may include symbolic references. 

We initially present the formalization of these expressions by referring to the tiny language without inheritance, and then discuss the handling of 
inheritance at the end of this section. The complete formalization of the \techniqueAcronym algorithm is provided in the Appendix.

\paragraph{Getfield expressions}
Figure~\ref{fig:semantics:getters} presents the formalization of getfield expressions, i.e., expressions in the form $\sigma.f$, being $\sigma$ a  reference value, and $f$ a field declared for a given class. As a result of evaluation, $\sigma$ could be a concrete reference, a plain symbolic reference, or a symbolic expression over reference values. In the former case the evaluation falls back to concrete execution. The figure shows the other two cases, indicated as rules Getfield~(c1) and Getfield~(c2), respectively. Getfield~(r1) and Getfield~(r2) formalize the related refinement steps. 

Rule Getfield~(c1) considers the case where $\sigma$ is a symbolic reference $Y$. The rule may fire only if the current configuration was suitably refined earlier in the symbolic execution: A symbolic object for $Y$ must exist in the heap ($Y \in \mathrm{dom}(H)$), and the the field $f$ must have some (possibly symbolic) value ($H(Y)(f) \neq \bot$). If such conditions are met, the value is read from the object field $H(Y)(f)$ leading to the transition \(H~\Sigma~Y.f \rightarrow H~\Sigma~H(Y)(f)\).

The refinement steps, Getfield~(r1) and Getfield~(r2), handle the assumptions needed to bind a symbolic reference $Y$ being accessed to a properly initialized object in the initial heap. In \techniqueAcronym, each symbolic reference is initially represented as a free symbolic value, which may not necessarily be bound to any object in the initial heap. In fact, the actual value of an input reference could be null in the initial state. However, when we assume the successful reduction of a getfield expression, we shall assume that the corresponding reference was bound to a \emph{symbolic object} in the initial heap. The refinement steps materialize this initial object, along with the assumption on the reference to be not null.

In detail, the rules  Getfield~(r1) and Getfield~(r2)  specify the refinements to associate the reference $Y$ to a symbolic object, and the field $Y.f$ to a value, respectively, in the cases when those initializations are needed.
Rule Getfield~(r1) handles the situation where a symbolic reference has no corresponding symbolic object in the heap yet---i.e., is \emph{unbound}. The refinement step adds a symbolic object in which all fields are not yet initialized (indicated as the value $\bot$ of the fields), binds $Y$ to the new object, and suitably updates the path condition by adding the assumption that $Y\neq \code{null}$. 

Rule Getfield~(r2) handles the access to an uninitialized field ($H(Y)(f) = \bot$) by initializing it with either a fresh symbolic value $Z$ (if $Y$ denotes a fresh object) or the value of the same field of another, type-compatible symbolic object (if $Y$ is alias of this object). Differently from lazy initialization, where all these alternative initializations for $Y$ would lead to branching the current symbolic state, in \techniqueAcronym they are encoded within the symbolic value of the field, by means of an ite-expression, indicated in the refinement rule by $assume(H, Y, f, Z)$. This predicate represents the possible alias relations between $Y.f$ and the field $Y'.f$ of other already bound and compatible objects, formally: 
\begin{multline*}    
assume(H, Y, f, Z) = \code{ite(} Y = Y_1, H(Y_1)(f), \ldots, \\
\code{ite(} Y = Y_m, H(Y_m)(f), Z \code{)} \ldots \code{)}
\end{multline*}
where $\{Y_1 \ldots Y_m\} = \{ Y' \in \mathrm{dom}(H) \ | \ Y' \neq Y \, \land\, f \in \mathrm{dom}(H(Y')) \, \land \,  H(Y')(f) \neq \bot \}$ is the \emph{alias set} of $Y$, and is composed by the set of all the bound symbolic references that point to a type-compatible object in which field $f$ already incurred initialization. The symbolic value $Z$ stands for the initial value of $Y.f$, which is a fresh symbolic value if $Y.f$ is not alias with respect to other symbolic references considered so far. The $assume(H, Y, f, Z)$ value that is assigned to the $Y.f$ field reflects the possible current value of the field: For all $Y_i$, if $Y$ aliases $Y_i$, then $Y.f$ has the same value stored in $Y_i.f$; otherwise, if $Y$ does not alias any object in its alias set, then $Y.f$ is the unmodified initial value $Z$.

Figure~\ref{fig:sampleprog1}, which we already discussed above, exemplifies the application of rules Getfield~(c1), Getfield~(r1) and Getfield~(r2) to a sample program. 

The rule Getfield~(c2) copes with $\code{ite(}\sigma, \sigma_1, \sigma_2\code{)}$ symbolic references. These symbolic references may emerge as expressions of alias conditions and are handled by reading the field $f$ from the objects denoted by the expressions $\sigma_1$ and $\sigma_2$, and returning the \code{ite} of the results. The effect on the heap and on the path condition is defined in terms of a suitable merging between the effects of reading the field of the objects referred to by $\sigma_1$ and $\sigma_2$, respectively. The resulting path condition, $\Sigma' = mergePC(\sigma, \Sigma'_1, \Sigma'_2)$ is obtained by conjoining all the clauses of $\Sigma'_1$ and $\Sigma'_2$, each guarded respectively by $\sigma$ and $\neg \sigma$. The definition of the merged heap, $H' = mergeHGf(H'_1, H'_2, f, \sigma)$ relies on the fact that $H'_1$ and $H'_2$ are mostly identical to $H$, differing only on the symbolic objects introduced by the Getfield~(r1) and Getfield~(r2) refinement rules. It is therefore sufficient to add all these objects (de facto identical in $H'_1$ and $H'_2$) to $H$ and obtain the merged heap $H'$.

\paragraph{Putfield expressions}
\begin{figure}[t!]
\begin{center}\scriptsize
\(
\inference[\scriptsize Putfield (c1):]{
Y \in \mathrm{dom}(H) \quad H_r = update(H, Y, f, \sigma')
}{
H \ \Sigma \ Y.f := \sigma' -> H_r[Y \mapsto H(Y)[f \mapsto \sigma']] \ \Sigma \ \sigma'
}
\)
\\~\\~\\
\(
\inference[\scriptsize Putfield (c2):]{
H \ \Sigma \ \sigma_1.f := \sigma' => H'_1 \ \Sigma'_1 \ \sigma' \quad H \ \Sigma \ \sigma_2.f := \sigma' => H'_2 \ \Sigma'_2 \  \sigma' \\
H' = mergeHPf(H'_1, H'_2, f, \sigma) \\
\Sigma' = mergePC(\sigma, \Sigma'_1, \Sigma'_2), mergeClauses(H'_1, H'_2, f, \sigma)
}{
H \ \Sigma \ \code{ite(}\sigma, \sigma_1, \sigma_2\code{)}.f := \sigma' -> H' \ \Sigma' \ \sigma'
}
\)
\\~\\~\\
\(
\inference[\scriptsize Putfield (r1):]{
Y \notin \mathrm{dom}(H) \\
H' = H, Y \mapsto (fields(class(f)) \mapsto \bot) \quad
\Sigma' = \Sigma, \neg Y = \code{null}
}{
H \ \Sigma \ Y.f := \sigma' ~> H' \ \Sigma' \ Y.f := \sigma'
}
\)
\end{center}
\caption{Refinement and computation transition relations for putfield expressions}
\label{fig:semantics:setters}
\vspace{-15pt}
\end{figure}

Figure~\ref{fig:semantics:setters} presents the formalization of field assignment expressions, $\sigma.f := \sigma'$, in the cases where $\sigma$ is either a plain symbolic reference $Y$ (rules Putfield~(c1) and Putfield~(r1)) or a symbolic expression over reference values $\code{ite}(\sigma,\sigma_1,\sigma_2)$ (rule Putfield~(c2)), respectively. Rule Putfield~(c1) is for updating a field of an object accessed via a symbolic reference $Y$. The update can be performed only if $Y$ is bound and the symbolic object has field $f$, which may require refinement Putfield~(r1) to bind $Y$ to an input object, in the same fashion as the refinement Getfield~(r1) described above. The effect of a field update does not necessarily affect only the symbolic object bound to $Y$: It may also affect all symbolic objects in the alias set of $Y$. Formally, the impact of the update on the alias set is described by the heap  $H_r = update(H, Y, f, \sigma')$, where
\begin{multline*}    
\mathrm{dom}(H_r)=\mathrm{dom}(H) \, \wedge \, \forall \,u' \in \mathrm{dom}(H), \, f' \in \mathrm{dom}(H)(Y') \,| \phantom{} \\ u' = Y' \, \land \, Y' \neq Y \, \land \, f' = f \, \land \, H(Y')(f) \neq \bot\\ \implies H_r(u')(f') = \code{ite(}Y = Y', \sigma', H(u')(f')\code{)} \, \wedge \\
\neg(u' = Y' \, \land \, Y' \neq Y \, \land \, f' = f \, \land \, H(Y')(f) \neq \bot)\\ \implies H_r(u')(f') = H(u')(f').
\end{multline*}    
Informally, if $Y'$ is a bound symbolic reference to a type-compatible object whose $f$ field is initialized, the field $Y'.f$ must also be updated to an ite reflecting the potential alias relation: If $Y$ aliases $Y'$, then $Y'.f$ also assumes the value assigned to $Y.f$, otherwise it retains its current value. 

Figure~\ref{fig:sampleprog2}, which we already discussed above, exemplifies the application of rules Putfield~(c1) and Putfield~(r1) to a sample program. 

The rule Putfield~(c2) handles the case where a symbolic reference $\code{ite(}\sigma, \sigma_1, \sigma_2\code{)}$ is used to refer to the object to be updated. Again, the effect on the heap and on the path condition is defined in terms of a suitable merging between the effects of updating the objects referred to by $\sigma_1$ and $\sigma_2$, respectively. While merging the path conditions is identical as for the case of the Getfield~(c2) rule, merging the heaps is more complex as the same symbolic object in the heaps to merge $H'_1$ and $H'_2$ may differ by the value at field $f$. In this case, the merged object must have the ite of these values as the resulting value at field $f$.

\paragraph{Conditional expressions}
The \techniqueAcronym semantics for conditional expressions that depend on conditions that predicate on references is straightforward. The only comparisons allowed between references are equality and equality with respect to \code{null}. This gives rise to three possible cases: Either two symbolic references are compared for equality, or a symbolic reference is compared for equality against a concrete reference, or a symbolic reference is compared for equality against \code{null}. In all cases, \techniqueAcronym will  simply conjoin the evaluated condition with the current path condition, either as is or after negating it (according to the taken branch), regardless of whether or not those symbolic references were bound to objects.

If the path condition includes other values that depend on alias conditions with the references that belong to the new clause, the new clause will impact on the solutions of those values, possibly causing contradictions (infeasible branches), as usual for branching statements in symbolic execution.  

\paragraph{Method invocations}
If we do not consider inheritance, handling method invocation expressions $Y.m(\ldots)$ requires only to update the path condition with the assumption on the value of $Y$ not to be null. Then the execution proceeds on the first instruction of $m$. 

\paragraph{Inheritance and polymorphism}
The \techniqueAcronym  approach to handling inheritance and polymorphic object's dereferencing is based on the classic approach of predicating type-order relations over the types of the input objects, according to the partial order between object types defined in the inheritance tree~\cite{kiasan:jase:2012,symbtypes:cc:2017}. 

While symbolically executing accesses to fields, \techniqueAcronym handles information on the type of the symbolic objects that are materialized. The complete forms of the rules Getfield~(r1) and Putfield~(r1), as reported in the external Appendix, add to the path condition assumptions with shape $Y \leq: c$, signifying that the symbolic object referred by $Y$ has a type that is ``\emph{more specialized than or the same as}'' class $c$ (i.e., the class that declares the field). Additional rules handle the case where the field being accessed belongs to a subtype of the currently assumed type for the symbolic object: Then, the object type is further refined in the path condition, and the missing fields are added to the symbolic object. 

The semantics of method invocations is the most impacted. With inheritance and polymorphism, a method invocation executed on a symbolic object generates a separate trace for each possible implementation of the method. This is consistent with the path optimality aim of \techniqueAcronym, since each trace  executes a distinct program path, which flows through a distinct implementation of the target method. In each trace the path condition is predicated by the assumption that the type of the receiver object is more specialized than the class that specifies the  method's implementation and is not more specialized than the subclasses (if any) that override the implementation. Notably, some assumptions done upon calling subsequent methods along the same path can result in contradictions on the type assumptions, when the call of the former method assumed a subtype that is incompatible with the latter method.

An in-depth presentation of symbolic execution with symbolic object types is out of the scope of this paper.
We refer readers to previous work~\cite{kiasan:jase:2012,symbtypes:cc:2017}.

%% file: experiments.tex
\section{Evaluation} \label{sec:experiments}
We implemented a prototype of a \techniqueAcronym-based symbolic executor that is written in Java and integrates with the Z3 SMT solver~\cite{de_moura_z3_2008}. The implementation of the prototype is mostly based on the code of the JBSE~\cite{braione_jbse_2016} symbolic executor, and it is able to analyze binary programs in Java bytecode. 
We used the prototype to experience the \techniqueAcronym algorithm against a set of subject programs, aiming to a proof-of-concept of the \techniqueAcronym idea.

\subsection{Research Questions}

We aimed to answer the following research questions:

\begin{description}
    \item[RQ1] To what extent does \techniqueAcronym fork less analysis traces than symbolic execution equipped with classic lazy initialization?
        
    \item[RQ2] Can \techniqueAcronym make symbolic execution incur less constraint solving load?

    \item[RQ3] Can \techniqueAcronym improve  efficiency in symbolic-execution-based test generation?
    
\end{description}

The research question RQ1 addresses the main hypothesis of this paper, that is, that 
\techniqueAcronym is effective for mitigating the path explosion issues of symbolic execution of heap manipulating programs.
As we discussed in Section~\ref{sec:language}, \techniqueAcronym enjoys the property of forking exactly an analysis trace for each analyzed program path.
We answer RQ1 by measuring the extent of the difference in the number of analysis traces that symbolic execution  forks when working either according to \techniqueAcronym or based on classic lazy initialization, respectively.

Following up, we quantify to what extent forking a path-optimal number of analysis traces can improve the efficiency of software engineering tools that rely on symbolic execution. 
We look at the efficiency improvements from a twofold angle.
On one hand, RQ2 investigates the extent to which a reduced number of analysis traces  maps to differences in the constraint solving load. We answer RQ2 by comparing the number of queries that symbolic execution issues for constraint solving and the time spent for solving those constraints, either with \techniqueAcronym or lazy initialization, respectively.

On the other hand, RQ3 widens the scope of our efficiency evaluation, by considering 
also the costs of post-processing the results of symbolic execution, as  tools based on symbolic execution must generally do, in order to address practical software engineering tasks, e.g., test generation.
In fact, the number of symbolic traces affects the amount of cognitive load for a tool to exploit the results of symbolic execution. 
For example, when using symbolic execution for path-driven test generation (e.g., as in dynamic symbolic execution~\cite{godefroid:dart:pldi:2005,sen:cute:esec:2005}), the number of analysis traces corresponds to the number of test cases to be instantiated and run during the  test generation process. The higher the number of symbolic traces, the higher the number of test cases to be instantiated and run, the higher the computation costs paid by the test generator (in addition to the cost of the symbolic analysis itself) for synthesizing the test cases in executable format, execute them, possibly store them, and interpret the results of each test execution.
We answer RQ3 by specifically focusing on the use case of symbolic execution for test generation,  comparing 
the overall costs of both executing symbolic execution and rendering test cases from 
the solutions computed for the program paths, either with \techniqueAcronym or lazy initialization, respectively.

\begin{table*}[ht!]
\caption{Results of \techniqueAcronym and JBSE  for the classes in the benchmark} \label{tab:results}
    \centering
    \scriptsize
    \begin{tabular}{l|rr|rr|rr|rr}
    %\hline
    & \multicolumn{2}{c|}{\#traces} & \multicolumn{2}{c|}{\#tokens}  & \multicolumn{2}{c|}{query time (sec)} & \multicolumn{2}{c}{test time (sec)}\\
    %\hline
            Subject & POSE & JBSE & POSE & JBSE & POSE & JBSE & POSE & JBSE\\\hline
        org.apache.bcel.generic.InstructionList & 18,793 & 61,404,980 & 37,355,572 & 385,255,412 & 2,090 & 105,195 & \bf 16,939 & $>$36,000 \\ 
        com.google.gson.internal.LinkedHashTreeMap & 13 & 3,363,888 & 5,691 & 1,177,354 & $<$1 & 235 & \bf 196 & $>$36,000 \\ 
        org.apache.commons.jxpath.ri.compiler.Path & 38 & 2,513,479 & 5,580 & 192,712,023 & $<$1 & 87,469 & \bf 203 & $>$36,000 \\ 
        org.apache.bcel.verifier.structurals.InstConstraintVisitor & 2,093 & 1,608,780 & 1,604,751 & 159,094,514 & 426 & 49,806 & \bf 7,164 & $>$36,000 \\ 
        org.la4j.matrix.dense.Basic1DMatrix & 1,533 & 1,367,819 & 1,415,265 & 81,278,176 & 9,067 & 15,047 & \bf 9,956 & %exp.
        $>$36,000 \\ 
        org.apache.bcel.generic.ConstantPoolGen & 227 & 1,333,160 & 432,833 & 277,680,886 & $<$1 & 209,260 & \bf 652 & $>$36,000 \\ 
        org.apache.bcel.generic.BranchInstruction & 229 & 1,233,323 & 2,919,833 & 72,661,571 & 518 & 10,937 & \bf 1,544 & %exp.
        $>$36,000 \\ 
        org.la4j.Matrix & 3,974 & 990,196 & 5,530,188 & 58,818,758 & 17,085 & 7,730 & \bf 20,478 & %exp. 
        $>$36,000 \\ 
        org.apache.commons.jxpath.JXPathContext & 79 & 956,887 & 222,459 & 1,433,926 & $<$1 & 43 & \bf 468 & $>$36,000 \\ 
        com.google.gson.JsonPrimitive & 85 & 941,686 & 15,856 & 76,468,387 & $<$1 & 14,651 & \bf 385 & %exp.
        $>$36,000 \\ 
        org.apache.bcel.verifier.structurals.LocalVariables & 12 & 282,895 & 3,446 & 125,280,089 & $<$1 & 51,744 & \bf 134 & $>$36,000 \\ 
        org.apache.bcel.classfile.StackMapEntry & 2,037 & 243,834 & 1,177,182 & 122,070,232 & 6,989 & 47,852 & \bf 7,444 & $>$36,000 \\ 
        org.la4j.matrix.sparse.CCSMatrix & 2,087 & 138,253 & 751,634 & 11,364,137 & 4,481 & 804 & \bf 4,871 & $>$36,000 \\ 
        org.apache.commons.jxpath.ri.parser.XPathParserTokenManager & 217 & 132,213 & 1,651,105 & 7,719,802 & 1,919 & 1,317 & \bf 1,873 & 20,309 \\ 
        
        org.la4j.matrix.sparse.CRSMatrix & 2,074 & 133,102 & 751,634 & 11,059,959 & 4,265 & 1,161 & \bf 5,118 & $>$36,000 \\ 
        com.google.re2j.Regexp & 33 & 113,775 & 5,610 & 11,777,518 & $<$1 & 11,868 & \bf 58 & $>$36,000 \\ 
        org.freehep.math.minuit.MnUserParameterState & 217 & 99,826 & 93,762 & 8,248,649 & 278 & 643 & \bf 1,277 & 8,629 \\ 
        org.la4j.vector.sparse.CompressedVector & 505 & 85,457 & 69,398 & 8,343,084 & 74 & 23,596 & \bf 258 & $>$36,000 \\ 
        org.freehep.math.minuit.MnAlgebraicSymMatrix & 46 & 80,203 & 21,176 & 5,369,421 & 69 & 746 & \bf 213 & 5,497 \\ 
        org.apache.commons.jxpath.ri.compiler.Step & 47 & 71,493 & 4,808 & 52,826,289 & $<$1 & 21,883 & \bf 57 & $>$36,000 \\ 
        
        org.la4j.matrix.dense.Basic2DMatrix & 1,480 & 72,549 & 715,700 & 4,885,025 & 4,640 & 1,683 & \bf 4,163 & 17,320 \\ 
            
        org.la4j.linear.ForwardBackSubstitutionSolver & 34 & 65,572 & 32,383 & 3,844,586 & 134 & 376 & \bf 177 & 10,438 \\ 
        
        org.la4j.linear.GaussianSolver& 913 & 53,473 & 898,386 & 3,131,321 & 6,293 & 774 & \bf 5,094 & 10,429 \\ 
        org.apache.commons.jxpath.ri.compiler.CoreOperationCompare & 42 & 24,326 & 10,452 & 651,350 & $<$1 & 319 & \bf 146 & $>$36,000 \\ 
        org.apache.commons.imaging.formats.tiff.TiffField & 235 & 13,743 & 520,140 & 7,288,795 & 516 & 895 & \bf 923 & 2,482 \\ 
        org.apache.commons.jxpath.ri.model.beans.PropertyIterator & 12 & 6,326 & 2,414 & 226 & $<$1 & $<$1 & \bf 90 & 9,201 \\ 
        okhttp3.HttpUrl & 27 & 6,332 & 15,346 & 24,661,699 & $<$1 & 1,053 & \bf 215 & 8,810 \\ 
        com.google.re2j.Machine & 921 & 4,546 & 56,389 & 203,369 & 7 & 92 & \bf 453 & 1,068 \\ 
        okhttp3.internal.tls.OkHostnameVerifier & 3 & 2,862 & 1,171 & 1,097,254 & $<$1 & 724 & \bf 42 & 1,766 \\ 
        org.freehep.math.minuit.MnHesse & 49 & 2,848 & 9,396 & 28,331,935 & $<$1 & 40,087 & \bf 326 & %56,037
        $>$36,000\\ 
        com.google.gson.GsonBuilder & 66 & 1,024 & 16,087 & 73,168 & $<$1 & 11 & 616 & \bf 168 \\ 
        org.freehep.math.minuit.MnUserTransformation & 89 & 642 & 23,085 & 18,344 & $<$1 & $<$1 & 898 & \bf 175 \\ 
        okhttp3.ConnectionSpec & 21 & 517 & 2,860 & 620,808 & $<$1 & 316 & \bf 38 & 674 \\ 
        okhttp3.Cookie & 282 & 599 & 1,511,800 & 654,915 & 3 & 328 & 669 & \bf 460 \\ 
        com.google.gson.internal.LinkedTreeMap & 29 & 260 & 6,836 & 374 & $<$1 & $<$1 & 134 & \bf 28 \\ 
        org.freehep.math.minuit.MnMinos & 66 & 252 & 12,843 & 2,634 & $<$1 & $<$1 &300 &\bf  54 \\ 
        com.google.gson.internal.Excluder & 11 & 168 & 3,874 & 72,732 & $<$1 & 10 & 84 & \bf 40 \\ 
        com.google.gson.internal.bind.JsonTreeReader & 44 & 194 & 22,795 & 10,790 & $<$1 & $<$1 & 90 & \bf 15 \\ 
        okhttp3.internal.platform.AndroidPlatform & 41 & 155 & 7,220 & 2,798 & 2 & $<$1 & 108 & \bf 26 \\ 
        org.freehep.math.minuit.MnFunctionCross & 14 & 88 & 1,602 & 2,907 & $<$1 & $<$1 & 62 & \bf 10 \\ 
        org.apache.commons.imaging.formats.bmp.BmpImageParser & 335 & 374 & 2,867,111 & 2,095,549 & 51 & 18 & 1,289 & \bf 92 \\ 
        com.google.gson.reflect.TypeToken & 19 & 46 & 3,252 & 112 & $<$1 & $<$1 & 55 & \bf  9 \\ 
        com.google.re2j.RE2 & 41 & 58 & 25,329 & 598 & $<$1 & $<$1 & 168 & \bf 26 \\ 
        org.apache.commons.jxpath.util.BasicTypeConverter & 21 & 28 & 4,318 & 196 & $<$1 & $<$1 & 97 & \bf 14 \\ 
        org.apache.commons.imaging.formats.tiff.write.TiffImageWriterBase & 16 & 21 & 2,715 & 112 & $<$1 & $<$1 & 58 & \bf 12 \\ 
        org.freehep.math.minuit.SimplexBuilder & 7 & 12 & 1,121 & 106 & $<$1 & $<$1 & 17 & \bf 3 \\ 
        org.apache.bcel.verifier.structurals.Subroutines & 6 & 8 & 1,707 & 84 & $<$1 & $<$1 & 41 & \bf 6 \\ 
        com.google.gson.internal.bind.ReflectiveTypeAdapterFactory & 2 & 3 & 652 & 28 & $<$1 & $<$1 & 13 & \bf 3 \\ 
        org.freehep.math.minuit.MnPlot & 6 & 6 & 2,086 & 112 & $<$1 & $<$1 & 46 & \bf 8 \\ 
        com.google.gson.stream.JsonReader & 4 & 4 & 627 & 103 & $<$1 & $<$1 & 13 & \bf 2 \\ 
        com.google.re2j.CharClass & 4 & 4 & 676 & 140 & 1 & $<$1 & 11 &\bf 2 \\ 
        okhttp3.internal.tls.DistinguishedNameParser & 3 & 3 & 4,458 & 67 & $<$1 & $<$1 & 12 &\bf 3 \\ 
        org.la4j.decomposition.EigenDecompositor & 3 & 3 & 650 & 88 & $<$1 & $<$1 & 14 &\bf 5 \\ 
        org.la4j.decomposition.SingularValueDecompositor & 3 & 3 & 1,092 & 56 & $<$1 & $<$1 & 27 &  \bf 6 \\ 
        org.apache.commons.imaging.common.RationalNumber & 2 & 2 & 1,032 & 56 & $<$1 & $<$1 & 22 & \bf 6 \\ 
    \end{tabular}\vspace{-15pt}
\end{table*}

\subsection{Experimental Setting} \label{ssec:experiments-setting}

To evaluate the effectiveness of symbolic execution using our proposed algorithm, \techniqueAcronym, we conducted a series of experiments on a benchmark of Java classes derived from the SBST Java  testing tool competition~\cite{panichella:comp:sbst:2017}. This benchmark includes 69 Java classes, which were originally selected to assess the capabilities of various test generation tools, including test generators based on symbolic execution. 

To compare \techniqueAcronym with the classical lazy initialization approach~\cite{khurshid:tacas:2003}, we executed each subject program using both our \techniqueAcronym prototype and the symbolic executor JBSE~\cite{braione_jbse_2016}, which implements lazy initialization for symbolic references. Each subject program consists of a Java class composed of multiple methods. We executed both tools on all methods of all classes and aggregated the results at the class level.

To prevent non-terminating executions caused by loops or recursive calls, we bounded the analysis of the program paths of each method to a maximum of 80 nested method calls and 150 loop iterations. Additionally, we enforced a time limit of 6 hours per method. Both \techniqueAcronym and JBSE employed a deterministic path selection strategy based on depth-first path traversal, which ensures consistent replicability of all tool runs.

To ensure a fair comparison, we post-processed the results to guarantee that both tools analyzed equivalent state spaces. Specifically, we retained only the results of i)~those methods for which both \techniqueAcronym and JBSE completed execution within the 6-hour time limit, and ii)~those methods for which \techniqueAcronym terminated and JBSE did not, but JBSE analyzed more execution traces than \techniqueAcronym. 

In the former case, the termination of both tools ensured that they both explored the same bounded state space, even if they used different path selection strategies. In the latter case, we conservatively retained the JBSE results, as they suffice for concluding (RQ1 and RQ3) that JBSE requires the analysis of more execution traces than \techniqueAcronym for those methods. Indeed the cut part of the analysis that JBSE would have done after the allowed 6 hours would just have just strengthened our conclusions. Furthermore, as expected, out of the methods for which JBSE terminated and \techniqueAcronym did not, there was no method for which \techniqueAcronym analyzed more execution traces than JBSE. 

In all other cases in which either tool did not terminate in the allowed 6 hours, the data are inconclusive on whether the observed differences can be due to the \techniqueAcronym prototype having explored a larger state-space portion than JBSE, or the vice-versa; thus we filtered out the results of those samples (110 methods), not to jeopardise the interpretability of the experiments.  
For the same reason, we filtered out 14 methods for which the \techniqueAcronym prototype computed relevant sets of analysis traces related to execution cases not yet supported by JBSE, namely, reflective executions that access objects of type java.lang.Class, executions of static class initializers, division-by-zero exceptions, and satisfiability decisions that depend on integer overflows.\footnote{The \techniqueAcronym prototype encodes path conditions with bit-vectors and can identify formulas satisfiable in case of  integer overflows, while JBSE relies on standard arithmetics over integers and cannot make those conclusions.} Also in these cases the observations cannot be interpreted because the \techniqueAcronym prototype arguably analyses a larger state-space portion than JBSE. In summary, of the 816 methods in the classes of the benchmark, the results in Table~\ref{tab:results} include the data of 692 methods, aggregated by their corresponding (55 out of 69) classes. 

\subsection{Results} 
\subsubsection{RQ1: Number of Analysis Traces}

\begin{figure}[t!]
  \begin{tabular}{cc}
  \includegraphics[width=.47\columnwidth]{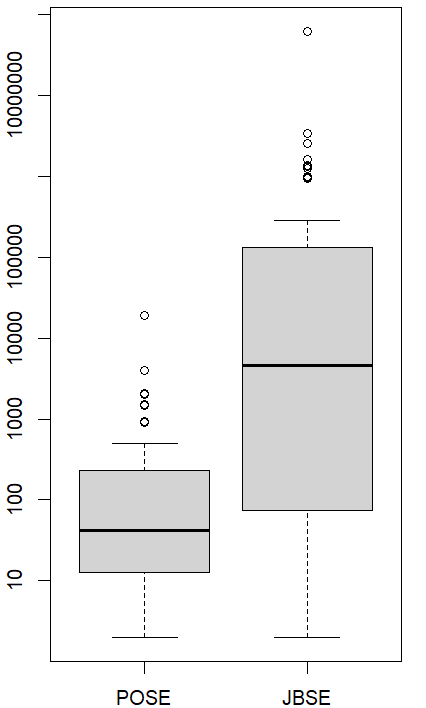}
  & \includegraphics[width=.47\columnwidth]{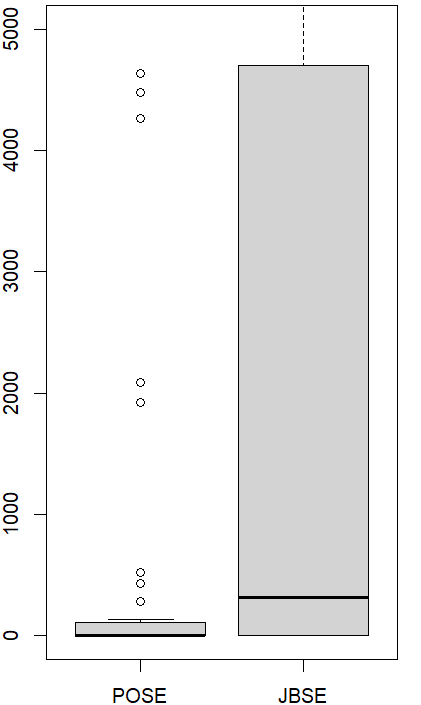}\\
  \footnotesize
  (a) \#traces &\footnotesize (b) query time (sec)
  \end{tabular}
  \footnotesize ~\\($^*$) In  boxplots (b) the y-axis is excerpted in the range (0, 5,000). 
  \caption{Distribution statistics of  \#traces and query time (\S Table~\ref{tab:results})} \label{fig:boxplots}\vspace{-15pt}
\end{figure}

Table~\ref{tab:results}, columns \emph{\#traces}
report the number of analysis traces forked during symbolic execution, 
with \techniqueAcronym and JBSE, respectively, for each considered class, in order of  decreasing difference of the numbers of traces in each row. 

The data in the table provide strong evidence in support of our claim that the path explosion incurred when using lazy initialization (with JBSE) can grow up to massive amounts, 
up to several orders of magnitude higher than the optimal number of program paths analyzed with \techniqueAcronym.
For instance, for the 2 subjects reported at the second and the third row of the table, \techniqueAcronym analyzed 5-orders-of-magnitude less  program paths than the number of analysis traces unfolded with JBSE through lazy initialization. 
The boxplots in Figure~\ref{fig:boxplots}~(a) visualize the consistently large extent of difference (reported with logarithmic scale)  in the number of traces, across the subjects.

\subsubsection{RQ2: Constraint Solving Load}

Table~\ref{tab:results}, columns \emph{\#tokens} and \emph{\#query time} report on the
queries sent to the constraint solver in each analysis session. The number of tokens measures the cumulative size of all formulas (i.e., the total  number of variables, constants and operators therein) sent as solver  queries. 
Counting the tokens captures 
the solver load better than the raw number of queries, as by-design \techniqueAcronym leads to submitting less queries, but also larger formulas (which include ite-expressions) per query, than JBSE. The 
query time reports the time taken for solving all  submitted queries. 

As expected, the data in the table pinpoint  cases in which \techniqueAcronym analyzes less traces than JBSE, but still it submits
cumulatively larger queries and takes more solving time than JBSE. Nonetheless, the boxplots in Figure~\ref{fig:boxplots}~(b) indicate that \techniqueAcronym brings significant gains on the solving time in most cases.
The Wilcoxon signed-rank test (paired) supports that the query time of \techniqueAcronym is significantly lower (p-value 0.0014) than in the case of JBSE through lazy intialization.

\subsubsection{RQ3: Test Generation Efficiency}

Table~\ref{tab:results}, columns \emph{test time} report the time to generate test cases with \techniqueAcronym
and JBSE, respectively. We recall that these experiments mimic the behavior of test generators that execute symbolic execution, but also post-process results of symbolic execution, using the solutions of the feasible program paths to synthesize and store test cases. We set  the time budget of the test generators to a maximum of 10 hours (36,000 seconds) for each class.

For the subjects listed in the highest part of the table, down to subject \textit{org.freehep.math.minuit.MnHesse}, the experimental data are consistently in favor of \techniqueAcronym. 
For these subjects, the results evidently demonstrate the beneficial impact of \techniqueAcronym that makes the test generation tool compute a path-optimal number of test cases. 
The test time of  \techniqueAcronym remains under control, consisting of  minutes in most of the cases, while  JBSE takes several hours or more than 10 hours most often.

On the other hand, the data are in favour of JBSE for  24 subjects in the lowest part of the table. Our inspection of the results indicated that, for these subjects, the \techniqueAcronym prototype pays the overhead of manipulating larger formulas than JBSE and the lower maturity of its implementation, with respect to a highly optimized tool such as JSBE. In these subjects, the overhead of \techniqueAcronym compensates for the gain of handling a (not enough) lower  number of traces than JBSE.

Even  acknowledging the overhead of \techniqueAcronym, our data testify that practical tools based on symbolic execution may strongly call for the path-optimal outcomes of  \techniqueAcronym.

\subsection{Threats to Validity} 

We did not engage in a direct comparison between \techniqueAcronym and lazier\# initialization~\cite{kiasan:jase:2012}, as there is no publicly accessible symbolic executor that implements it. In Section~\ref{sec:motivating} we showed that lazier\# initialization easily falls back to lazy initialization when programs access the data stored in the input fields, as it occurs in the subject programs that we considered. We expect that the results that we computed with JBSE for the considered subjects may extend with minor differences to the case of  generalized symbolic execution based on lazier\# initialization. 

\TechniqueAcronym has been designed to unfold exactly the same set of symbolic-reference assumptions of lazy initialization, though modeling those assumptions with comprehensive ite-expressions rather than spawning a separate symbolic-execution trace for each assumption. We leave for future work the formal proof that \techniqueAcronym is sound and complete with respect to the traditional embodiment of generalized symbolic execution. 
Currently we manually crosschecked for a sample of the methods in the experimental benchmark that 
the numbers of analysis traces computed with \techniqueAcronym corresponded to the exact number of program paths as expected.

The \techniqueAcronym prototype and JBSE differ in  
several ways with respect to their internal implementations.
To avoid biases, on one hand, we  filtered out the methods (\S\ref{ssec:experiments-setting}) for which the \techniqueAcronym prototype handled execution cases that JBSE does not support, namely, reflective executions that access objects of type java.lang.Class, executions of static class initializers, division-by-zero exceptions, and satisfiability decisions that depend on integer overflows. 

With respect to external validity, we experimented with a benchmark of programs  selected from the well-known SBST Java testing tool competition~\cite{panichella:comp:sbst:2017}, consisting of a heterogeneous set of applications, namely,  a bytecode instrumentation framework, a linear algebra library, an XML parser, an object serializer library, a framework for image manipulation, a regular expression library, and an HTTP client. 
As this benchmark was originally selected to assess the capabilities of various test generation tools, including test generators based on symbolic execution, we believe that this benchmark is representative of possible uses of symbolic execution in software engineering. 
Nonetheless, we are aware that the results of a single set of experiments cannot be directly generalized. We aim to replicate our experiments with further subjects, including programming languages other than Java, in the future.

%% file: related.tex
\section{Related work} \label{sec:related}
We already commented in Section~\ref{sec:motivating} on the relations between \techniqueAcronym and the previous work on generalized symbolic execution based on \emph{lazy initialization}~\cite{khurshid:tacas:2003,deng_bogor_kiasan_2006,laziersharp:sefm:2007,kiasan:jase:2012}. Lazy initialization 
straightforwardly augments classic symbolic execution for handling heap manipulating programs, but can heavily exacerbate the path explosion issues of symbolic execution. We provided empirical evidence of this phenomenon in Section~\ref{sec:experiments}. 

\TechniqueAcronym can be related with research on using \emph{state merging} to alleviate the path explosion issues of symbolic execution~\cite{constructing:ijpp:2005,babic:phd:2008,efficient:date:2008,statejoining:rv:2009,statemerging:pldi:2012}. Most state-merging techniques work on the traces spawned by the trace-unfolding algorithm, grouping together the symbolic states that, at given program points, match with each other according to possible merging criteria, reducing the amount of analysis traces analyzed thereon.
\TechniqueAcronym embraces technical means on the other side of the spectrum: It does not spawn separate symbolic traces that get eventually merged, rather it synthesizes suitably merged states up front. 

Sen et al.\ and Hillery et al.\ 
exploited guarded value sets to make lazy initialization enumerate multiple values~\cite{sen:multise:fse:2015,hillery:summaries:vmcai:2016}, rather than multiple traces, thus achieving similar effects as \techniqueAcronym. The \techniqueAcronym approach relies on ite-expressions, which may scale more effectively than guarded values sets when composing the symbolic values of multiple  variables that participate in the same assignment and comparison statements. As a matter of facts, the above seminal papers reported only experiments with a small number of recursive data structures and those approaches have not been further developed in the last ten years, which can be due to the low scalability. 
In contrast, in our empirical assessment,  \techniqueAcronym successfully handled a large benchmark of 692 Java methods belonging to various software projects.

%% file: conclusions.tex
\section{Conclusions} \label{sec:conclusions}
Current approaches to symbolic execution of heap-manipulating programs exacerbate the path explosion problem as they fork many more execution traces than the number of control flow paths in the program under analysis. This paper presented a novel approach, \techniqueAcronym, that keeps the number of program traces explored by symbolic execution isomorphic to the number of analyzed program paths. Our experiments, executed with a prototype implementation of \techniqueAcronym,  provide initial empirical support for our research hypothesis that 
\techniqueAcronym can  crucially improve the efficiency of symbolic execution for heap-manipulating programs.
We are currently working to 
extend the experiments, and to investigate how to integrate \techniqueAcronym with state merging techniques.

\section{Data availability}
The \techniqueAcronym prototype and a replication package are available 
at \url{https://github.com/pietrobraione/SANER2026}

%% file: formal_long.tex
\section{Full operational semantics for \techniqueAcronym}
\label{sec:formalization}

%Appendix A is provided in a separate version of the manuscript submitted as supplementary material.

\input{formal_long_text}

%% file: formal_long_text.tex
\begin{figure*}[tp]
\begin{align*}
& \text{Programs} && P && ::= && \rep{C}\ e \\
& \text{Classes} && C && ::= && \code{class} \ c : c \ \{ \rep{t \ f}; \rep{D} \} \\
& \text{Methods} && D && ::= && t \ m(\rep{t \ x}) := e \\
& \text{Program expressions} && e && ::= && x \ | \ v \ | \ \code{new} \ c \ | \ e.f \ | \ e.f := e  \ | \ \code{let} \ x := e \ \code{in} \ e \\
& && && && | \ e \ op \ e \ | \ op \ e \ | \ \code{if} \ e \ e \ e  \ | \ e.m(\rep{e})  \\
& \text{Types} && t && ::= && c \ | \ \code{bool} \ | \ \code{int} \\
& \text{Values} && v && ::= && p \ | \ u  \\
& \text{Primitive values} && p && ::= && r \ | \ X \\
& \text{Reference values} && u && ::= && \code{null} \ | \ l \ | \ Y  \\
& \text{Literals} && r && ::= && b \ | \ n \\
& \text{Booleans} && b && ::= && \code{true} \ | \ \code{false} \\
& \text{Integers} && n && ::= && \ldots -1 \ | \ 0 \ | \ 1 \ldots \\
& \text{Operators} && op && ::= && + \ | \ = \ldots \\
& \text{Locations} && l \\
& \text{Identifiers} && \omit\rlap{$c, f, m, x, X, Y$} 
\end{align*}
\caption{Language syntax}
\label{fig:syntax-lang}
\end{figure*}

\begin{figure*}
\begin{align*}
& \text{Configurations} && J && ::= && H \ \Sigma \ \eta \\
& \text{Configuration expressions} && \eta && ::= && x \ | \ \sigma \ | \ \code{new} \ c \ | \ \eta.f \ | \ \eta.f := \eta  \ | \ \code{let} \ x := \eta \ \code{in} \ \eta \\
& && && && | \ \eta \ op \ \eta \ | \ op \ \eta \ | \ \code{if} \ \eta \ \eta \ \eta  \ | \ \eta.m(\rep{\eta})  \\
& \text{Heaps} && H && ::= && \rep{u \mapsto o}, \ u \neq \code{null} \\
& \text{Objects} && o && ::= && (\rep{f \mapsto \sigma})^c \ | \ (\rep{f \mapsto \sigma}) \\
& \text{Path conditions} && \Sigma && ::= && \emptyset \ | \ \Sigma, \sigma \\
& \text{Symbolic values} && \sigma && ::= && \bot \ | \ v \ | \ \sigma \ op\ \sigma \ | \ op\ \sigma \ | \ Y \leq : c \ | \ Y.f = Z  \ | \ \code{ite(} Y = Y, \sigma, \sigma \code{)} \\
& \text{Evaluation contexts} && E && ::= && - \ | \ E.f \ | \ E.f := \eta \ | \ \sigma.f := E \ | \ \code{let}\ x := E \ \code{in}\ \eta \\
& && && && | \ E \ op \ \eta \ | \ \sigma \ op \ E \ | \ op \ E \ | \ \code{if}\ E \ \eta \ \eta \ | \ E.m(\rep{\eta}) \\
& && && && | \ \sigma.m(\rep{\sigma}, E, \rep{\eta})
\end{align*}
\caption{Syntax of configurations and evaluation contexts}
\label{fig:syntax-config-contexts}
\end{figure*}

\begin{figure*}
\begin{center}    
\(
\inference[Step:]{
H \ \Sigma \ \eta ~>^* H' \ \Sigma' \ \eta \quad H' \ \Sigma' \ \eta -> H'' \ \Sigma'' \ \eta''
}
{
H \ \Sigma \ \eta => H'' \ \Sigma'' \ \eta''
}
\)
\end{center}
\caption{Small-step operational semantics}
\label{fig:semantics}
\end{figure*}

\begin{figure*}
\begin{center}    
\(
\inference[New:]{
fields(c) = \rep{f} \quad l \notin \mathrm{dom}(H)
}
{
H \ \Sigma \ \code{new} \ c -> H, l \mapsto (\rep{f \mapsto default})^c \ \Sigma \ l
}
\)
\\~\\~\\
\(
\inference[Getfield (c1):]{
Y \in \mathrm{dom}(H) \quad f \in \mathrm{dom}(H(Y)) \quad H(Y)(f) \neq \bot
}{
H \ \Sigma \ Y.f -> H \ \Sigma \ H(Y)(f)
}
\)
\\~\\~\\
\(
\inference[Getfield (c2):]{
H \ \Sigma \ \sigma_1.f => H'_1 \ \Sigma'_1 \ \sigma'_1 \quad H \ \Sigma \ \sigma_2.f => H'_2 \ \Sigma'_2 \ \sigma'_2 \\ H' = mergeHGf(H'_1, H'_2, f, \sigma) \quad \Sigma' = mergePC(\sigma, \Sigma'_1, \Sigma'_2)
}{
H \ \Sigma \ \code{ite(}\sigma, \sigma_1, \sigma_2\code{)}.f -> H' \ \Sigma' \ \code{ite(}\sigma, \sigma'_1, \sigma'_2 \code{)}
}
\)
\\~\\~\\
\(
\inference[Getfield (c3):]{
H \ \Sigma \ \sigma_1.f => H' \ \Sigma' \ \sigma'_1 \quad H \ \Sigma \ \sigma_2.f \not\Rightarrow\\ 
}{
H \ \Sigma \ \code{ite(}\sigma, \sigma_1, \sigma_2\code{)}.f -> H' \ \Sigma',\sigma \ \sigma'_1
}
\)
\\~\\~\\
\(
\inference[Getfield (c4):]{
H \ \Sigma \ \sigma_1.f \not\Rightarrow \quad H \ \Sigma \ \sigma_2.f => H' \ \Sigma' \ \sigma'_2\\ 
}{
H \ \Sigma \ \code{ite(}\sigma, \sigma_1, \sigma_2\code{)}.f -> H' \ \Sigma',\neg\sigma \ \sigma'_2
}
\)
\\~\\~\\
\(
\inference[Getfield (c5):]{
}{
H \ \Sigma \ l.f -> H \ \Sigma \ H(l)(f)
}
\)
\\~\\~\\
\(
\inference[Putfield (c1):]{
Y \in \mathrm{dom}(H) \quad f \in \mathrm{dom}(H(Y)) \quad H_r = update(H, Y, f, \sigma')
}{
H \ \Sigma \ Y.f := \sigma' -> H_r[Y \mapsto H(Y)[f \mapsto \sigma']] \ \Sigma \ \sigma'
}
\)
\\~\\~\\
\(
\inference[Putfield (c2):]{
H \ \Sigma \ \sigma_1.f := \sigma' => H'_1 \ \Sigma'_1 \ \sigma' \quad H \ \Sigma \ \sigma_2.f := \sigma' => H'_2 \ \Sigma'_2 \  \sigma' \\
H' = mergeHPf(H'_1, H'_2, f, \sigma) \\ \Sigma' = mergePC(\sigma, \Sigma'_1, \Sigma'_2), mergeClauses(H'_1, H'_2, f, \sigma)
}{
H \ \Sigma \ \code{ite(}\sigma, \sigma_1, \sigma_2\code{)}.f := \sigma' -> H' \ \Sigma' \ \sigma'
}
\)
\\~\\~\\
\(
\inference[Putfield (c3):]{
H \ \Sigma \ \sigma_1.f := \sigma' => H' \ \Sigma' \ \sigma' \quad H \ \Sigma \ \sigma_2.f := \sigma' \not\Rightarrow\\
}{
H \ \Sigma \ \code{ite(}\sigma, \sigma_1, \sigma_2\code{)}.f := \sigma' -> H' \ \Sigma',\sigma \ \sigma'
}
\)
\\~\\~\\
\(
\inference[Putfield (c4):]{
H \ \Sigma \ \sigma_1.f := \sigma' \not\Rightarrow \quad H \ \Sigma \ \sigma_2.f := \sigma' => H' \ \Sigma' \ \sigma' \\
}{
H \ \Sigma \ \code{ite(}\sigma, \sigma_1, \sigma_2\code{)}.f := \sigma' -> H' \ \Sigma',\neg\sigma \ \sigma'
}
\)
\\~\\~\\
\(
\inference[Putfield (c5):]{
}{
H \ \Sigma \ l.f := \sigma' -> H[l \mapsto H(l)[f \mapsto \sigma']] \ \Sigma \ \sigma'
}
\)
\\~\\~\\
\(
\inference[Let:]{
}{
H \ \Sigma \ \code{let}\ x := \sigma \ \code{in} \ e -> H \ \Sigma \ e[\sigma/x]
} \qquad 
\inference[Op (c1):]{
r_1 \ op \ r_2 = r
}{
H \ \Sigma \ r_1 \ op \ r_2 -> H \ \Sigma \ r
}
\)
\\~\\~\\
\(
\inference[Op (c2):]{
op \ r_1 = r
}{
H \ \Sigma \ op \ r_1 -> H \ \Sigma \ r
}
\qquad
\inference[Eq (c1):]{
}{
H \ \Sigma \ v = v -> H \ \Sigma \ \code{true}
}
\)
\\~\\~\\
\(
\inference[Eq (c2):]{
r_1 \neq r_2
}{
H \ \Sigma \ r_1 = r_2 -> H \ \Sigma \ \code{false}
} \qquad
\inference[Eq (c3):]{
l_1 \neq l_2
}{
H \ \Sigma \ l_1 = l_2 -> H \ \Sigma \ \code{false}
}
\)
\\~\\~\\
\(
\inference[Eq (c4):]{
}{
H \ \Sigma \ l = \code{null} -> H \ \Sigma \ \code{false}
}\qquad
\inference[Eq (c5):]{
}{
H \ \Sigma \ \code{null} = l -> H \ \Sigma \ \code{false}
}
\)
\\~\\~\\
\(
\inference[Eq (c6):]{
}{
H \ \Sigma \ Y = l -> H \ \Sigma \ \code{false}
}
\qquad
\inference[Eq (c7):]{
}{
H \ \Sigma \ l = Y -> H \ \Sigma \ \code{false}
}
\)
\\~\\~\\
%\(
% \[
% \inference[Instanceof 1:]{
% }{
% H \ \Sigma \ \code{null} \ \code{instanceof}\ c -> H \ \Sigma \ \code{false}
% }
% \]
% \[
% \inference[Instanceof 2:]{
% H(l) \leq: c
% }{
% H \ \Sigma \ l \ \code{instanceof}\ c -> H \ \Sigma \ \code{true}
% }
% \]
% \[
% \inference[Instanceof 3:]{
% \neg H(l) \leq: c
% }{
% H \ \Sigma \ l \ \code{instanceof}\ c -> H \ \Sigma \ \code{false}
% }
% \]
% \[
% \inference[Instanceof 4:]{
% }{
% H \ \Sigma \ Y \ \code{instanceof}\ c -> H \ \Sigma \ Y \leq : c
% }
% \]
% \[
% \inference[Instanceof 5:]{
% H \ \Sigma \ \sigma_1  \ \code{instanceof}\ c => H \ \Sigma \ \sigma'_1 \quad H \ \Sigma \ \sigma_2  \ \code{instanceof}\ c => H \ \Sigma \  \sigma'_2
% }{
% H \ \Sigma \ \code{ite(}\sigma, \sigma_1, \sigma_2\code{)} \ \code{instanceof}\ c -> H \ \Sigma \ \code{ite(}\sigma, \sigma'_1, \sigma'_2 \code{)}
% }
% \]
\end{center}
\caption{Small-step operational semantics, computation transition (pt.~1)}
\label{fig:semantics-computation-1}
\end{figure*}

\begin{figure*}
\begin{center}
\(
\inference[If (c1):]{
}{
H \ \Sigma \ \code{if}\ \code{true} \ e_1 \ e_2 -> H \ \Sigma \ e_1
}
\)
\\~\\~\\
\(
\inference[If (c2):]{
}{
H \ \Sigma \ \code{if}\ \code{false} \ e_1 \ e_2 -> H \ \Sigma \ e_2
}
\)
\\~\\~\\
\(
\inference[If (c3):]{
}{
H \ \Sigma \ \code{if}\ \sigma \ e_1 \ e_2 -> H \ \Sigma, \sigma \ e_1
} \qquad 
\inference[If (c4):]{
}{
H \ \Sigma \ \code{if}\ \sigma \ e_1 \ e_2 -> H \ \Sigma, \neg \sigma \ e_2
}
\)
\\~\\~\\
\(
\inference[Invoke (c1):]{
H(l) = o^c \quad recv(c, m) = c' \quad method(c', m) = t \ m \ (\rep{t \ x}) \ \{ e_m \}
}{
H \ \Sigma \ l.m(\rep{\sigma}) -> H \ \Sigma \ e_m[l/\code{this}][\rep{\sigma/x}]
}
\)
\\~\\~\\
\(
\inference[Invoke (c2):]{
method(c', m) = t \ m \ (\rep{t \ x}) \ \{ e_m \} \quad O = \{ c \ | \ overrides(c, m, c') \}
}{
H \ \Sigma \ Y.m(\rep{\sigma}) -> H \ \Sigma, \neg Y = null, Y \leq: c', \rep{\neg Y \leq: c}_{c \in O}  \ e_m[l/\code{this}][\rep{\sigma/x}]
}
\)
\\~\\~\\
\(
\inference[Invoke (c3):]{
H \ \Sigma \ \sigma_1.m(\rep{\sigma}) -> H \ \Sigma'_1 \ \eta'_1
}{
H \ \Sigma \ \code{ite(}\sigma, \sigma_1, \sigma_2\code{)}.m(\rep{\sigma}) -> H \ \Sigma'_1,\sigma \ \eta'_1
}
\)
\\~\\~\\
\(
\inference[Invoke (c4):]{
H \ \Sigma \ \sigma_2.m(\rep{\sigma}) -> H \ \Sigma'_2 \ \eta'_2
}{
H \ \Sigma \ \code{ite(}\sigma, \sigma_1, \sigma_2\code{)}.m(\rep{\sigma}) -> H \ \Sigma'_2,\neg \sigma \ \eta'_2
}
\)
\\~\\~\\
\(
\inference[Ctx (c):]{
H \ \Sigma \ \eta -> H' \ \Sigma' \ \eta'
}
{
H \ \Sigma \ E[\eta / -] -> H' \ \Sigma' \ E[\eta' / -]
}
\)
\end{center}
\caption{Small-step operational semantics, computation transition (pt.~2)}
\label{fig:semantics-computation-2}
\end{figure*}

\begin{figure*}
\begin{center}
\(
\inference[Getfield (r1):]{
Y \notin \mathrm{dom}(H) \quad fields(class(f)) = \rep{f'}
}{
H \ \Sigma \ Y.f ~> H, Y \mapsto (\rep{f' \mapsto \bot}) \ \Sigma, \neg Y = \code{null}, Y \leq: c \ Y.f
}
\)
\\~\\~\\
\(
\inference[Getfield (r1.b):]{
Y \in \mathrm{dom}(H) \quad f \notin \mathrm{dom}(H(Y)) \quad fields(class(f)) \setminus fields(H(Y)) = \rep{f'}
}{
H \ \Sigma \ Y.f ~> H[Y \mapsto (H(Y),\rep{f' \mapsto \bot})] \ \Sigma, Y \leq: c \ Y.f
}
\)
\\~\\~\\
\(
\inference[Getfield (r2):]{
Y \in \mathrm{dom}(H) \quad f \in \mathrm{dom}(H(Y)) \quad H(Y)(f) = \bot \quad Z\ \text{fresh}
}{
H \ \Sigma \ Y.f ~> H[Y \mapsto H(Y) [f \mapsto assume(H, Y, f, Z)]] \ \Sigma, Y.f = Z \ Y.f
}
\)
\\~\\~\\
\(
\inference[Putfield (r1):]{
Y \notin \mathrm{dom}(H) \quad fields(class(f)) = \rep{f'}
}{
H \ \Sigma \ Y.f := \sigma' ~> H, Y \mapsto (\rep{f' \mapsto \bot}) \ \Sigma, \neg Y = \code{null}, Y \leq: c \ Y.f := \sigma'
}
\)
\\~\\~\\
\(
\inference[Putfield (r1.b):]{
Y \in \mathrm{dom}(H) \quad f \notin \mathrm{dom}(H(Y)) \quad fields(class(f)) \setminus fields(H(Y)) = \rep{f'}
}{
H \ \Sigma \ Y.f := \sigma' ~> H[Y \mapsto (H(Y),\rep{f' \mapsto \bot})] \ \Sigma, Y \leq: c \ Y.f := \sigma'
}
\)
\\~\\~\\
\(
\inference[Ctx (r):]{
H \ \Sigma \ \eta ~> H' \ \Sigma' \ \eta'
}
{
H \ \Sigma \ E[\eta / -] ~> H' \ \Sigma' \ E[\eta' / -]
}
\)
\end{center}
\caption{Small-step operational semantics, refinement transition}
\label{fig:semantics-refinement}
\end{figure*}

In this Appendix we introduce an example imperative, object-oriented programming language and give operational semantics for \technique.
Figure~\ref{fig:syntax-lang} reports the syntax of the language. A program is a sequence of class declarations followed by an expression to be evaluated. Classes have exactly one superclass (there is no multiple inheritance nor interfaces) and multiple fields and methods. The body of a method is a single expression.

There are many types of expressions, but here we want focus on expressions that read the content of object fields (getfield expressions $e.f$), modify the content of object fields (putfield expressions $e.f := e$), conditional expressions $\code{if\ } e\ e\ e$, and method invocation expressions $e.m(\rep{e})$. These are the expressions whose semantics is more affected by the fact that their subexpressions may be symbolic.

Figure~\ref{fig:syntax-config-contexts} reports the syntax of the configurations and the evaluation contexts used in the operational semantics definition. The small-step operational semantics transition relation $=>$ is defined in Figure~\ref{fig:semantics} as a suitable composition of two distinct transition relations, the \emph{computation} transition relation $->$, and the (reflexive-transitive closure of the) \emph{refinement} transition relation $~>$. These relations are formally defined in Figure~\ref{fig:semantics-computation-1}, Figure~\ref{fig:semantics-computation-2} and Figure~\ref{fig:semantics-refinement}. The computation transition accounts for the steps that implement a computational step performed to progress the evaluation of the current expression in the program. The refinement transition adds information to the current configuration that is already implicit in it, without progressing the evaluation of the program. The added information may come in the shape of a heap object, or of a value in a field of an object.

In \techniqueAcronym each symbolic reference points to a \emph{symbolic object} in the initial heap. In the beginning of symbolic execution these objects do not actually exist in the initial heap, and are materialized on demand (through a suitable refinement step) when the corresponding symbolic reference is used to access the object. This is the only similarity of \techniqueAcronym with generalized symbolic execution. Differently from the latter, where an input reference is initialized to point to the object it may alias to, in \techniqueAcronym  every symbolic reference points to a distinct symbolic object, and the alternative effects of the alias relationship are encoded \emph{inside each object}, by means of if-then-else-values (ite-values, with shape $\code{ite}(\ldots)$). 

Let us consider the computation semantics of getfield expressions $e.f$, that read the content of field $f$ from a heap object denoted by expression $e$. Rule Getfield~(c5) considers the case where $e$ evaluates to the location $l$ of a \emph{concrete object}, i.e., of an object created after the start of symbolic execution by means of a \code{new} expression. In this case, the value is simply read from the object field. Rule Getfield~(c1) considers the case where $e$ evaluates to a symbol $Y$ with reference type. The rule may fire only if a symbolic object for $Y$ exists in the heap ($Y \in \mathrm{dom}(H)$), the field $f$ is present in the symbolic object ($f \in \mathrm{dom}(H(Y))$), and the slot corresponding to the field $f$ has been initialized with some value ($H(Y)(f) \neq \bot$). We will discuss all these conditions when explaining the refinement transitions. When all the conditions are met, the value is read from the object field, exactly as for concrete objects. Finally, rules Getfield~(c2)-(c4) cope with $\code{ite(}\sigma, \sigma_1, \sigma_2\code{)}$ symbolic references. These symbolic references may emerge as expressions of alias conditions and are handled by reading the field $f$ from the objects denoted by the expressions $\sigma_1$ and $\sigma_2$, and returning the ite-expression combining the results (rule Getfield~(c2)). The resulting path condition is obtained by suitably merging the path conditions obtained by evaluating $\sigma_1.f$ and $\sigma_2.f$, by means of the function $\Sigma' = mergePC(\sigma, \Sigma'_1, \Sigma'_2)$ defined as follows: If $\Sigma'_1 = \sigma_{11}, \ldots, \sigma_{1m}$ and $\Sigma'_2 = \sigma_{21}, \ldots, \sigma_{2w}$, then $\Sigma' = (\sigma \implies \sigma_{11}), \ldots, (\sigma \implies \sigma_{1m}), (\neg \sigma \implies \sigma_{21}), \ldots, (\neg \sigma \implies \sigma_{2w})$. Similarly, the resulting heap is obtained by merging the updates that the evaluations of $\sigma_1.f$ and $\sigma_2.f$ apply to the heap, by means of the function $H' = mergeHGf(H'_1, H'_2, f, \sigma)$ defined as follows: $\mathrm{dom}(H')=\mathrm{dom}(H'_1) \cup \mathrm{dom}(H'_2)$, and for all $u \in \mathrm{dom}(H')$ it is:

\begin{enumerate}
\item if $u \in \mathrm{dom}(H'_1)$ and $u \notin \mathrm{dom}(H'_2)$, then $H'(u)=H'_1(u)$;
\item if $u \in \mathrm{dom}(H'_2)$ and $u \notin \mathrm{dom}(H'_1)$, then $H'(u)=H'_2(u)$;
\item if $u \in \mathrm{dom}(H'_1)$, and $u \in \mathrm{dom}(H'_2)$, and $H'_1(u) = H'_2(u)$, then $H'(u) = H'_1(u) = H'_2(u)$;
\item if $u \in \mathrm{dom}(H'_1)$, and $u \in \mathrm{dom}(H'_2)$, and $H'_1(u) \neq H'_2(u)$, then $\mathrm{dom}(H'(u)) = \mathrm{dom}(H'_1(u)) \cup \mathrm{dom}(H'_2(u))$, and for all $f \in \mathrm{dom}(H'(u))$:
\begin{enumerate}
\item if $f \in \mathrm{dom}(H'_1(u))$ and $f \notin \mathrm{dom}(H'_2(u))$, then $H'(u)(f)=H'_1(u)(f)$;
\item if $f \in \mathrm{dom}(H'_2(u))$ and $f \notin \mathrm{dom}(H'_1(u))$, then $H'(u)(f)=H'_2(u)(f)$;
\item if $f \in \mathrm{dom}(H'_1(u))$, and $f \in \mathrm{dom}(H'_2(u))$, and $H'_1(u)(f) = H'_2(u)(f)$ then $H'(u)(f)=H'_1(u)(f)=H'_2(u)(f)$;
\item if $f \in \mathrm{dom}(H'_1(u))$, and $f \in \mathrm{dom}(H'_2(u))$, and $H'_2(u)(f) = \bot$ then $H'(u)(f)=H'_1(u)(f)$;
\item if $f \in \mathrm{dom}(H'_1(u))$, and $f \in \mathrm{dom}(H'_2(u))$, and $H'_1(u)(f) = \bot$ then $H'(u)(f)=H'_2(u)(f)$;
\item if $f \in \mathrm{dom}(H'_1(u))$, and $f \in \mathrm{dom}(H'_2(u))$, and $H'_1(u)(f) \neq H'_2(u)(f)$, and $H'_1(u)(f) \neq \bot$, and $H'_2(u)(f) \neq \bot$ then $H'(u)(f)=\code{ite(}\sigma, H'_1(u)(f), H'_2(u)(f)\code{)}$.
\end{enumerate}
%\item  if $H'_1(u) = H'_2(u)$ then $H'(u) = H'_1(u) = H'_2(u)$;
%\item if $H'_1(u) \neq H'_2(u)$, then $u$ is a symbolic reference input (that we indicate with $Y$), $H'_1(Y) = H'_2(Y)[f \mapsto Z]$, $H'_2(Y) = H'_1(Y)[f \mapsto \bot]$ or $H'_1(Y) = H'_2(Y)[f \mapsto \bot]$, $H'_2(Y) = H'_1(Y)[f \mapsto Z]$ for some $Z$: in this case it is $H'(Y)=H'_1(Y)[f \mapsto Z] = H'_2(Y)[f \mapsto Z]$;
%\item if $u \notin \mathrm{dom}(H'_2)$ and $H'_1(u)(f) = Z$, then $u$ is a symbolic reference input (that we indicate with $Y$), and $H'(Y)=H'_1(Y)$;
%\item if $u \notin \mathrm{dom}(H'_2)$ and $H'_2(u)(f) = Z$, then $u$ is a symbolic reference input (that we indicate with $Y$), and $H'(Y)=H'_2(Y)$.
\end{enumerate}

In the case one of the field access expression $\sigma_1.f$, $\sigma_2.f$ cannot evaluate, the ite-based getfield reduces to the other field access expression (rules Getfield~(c3) and (c4)).

Now we analyze the refinement semantics of getfield expressions. Rule Getfield~(r1) handles the situation where a symbolic reference has no corresponding symbolic object in the heap yet---i.e., is \emph{unbound}. For the sake of simplicity we suppose that each field $f$ is contained in exactly one class. If $c = class(f)$ is the class that contains the field $f$, then the refinement step adds a symbolic object of class $c$ to the heap, whose fields are all uninitialized (the function $fields(c)$ yields the set of all the fields declared in $c$ and in its superclasses), binds $Y$ to the new object, and updates the path condition to reflect the fact that field access has success ($\neg Y = \code{null}$) and that the symbolic object's type is a subclass of $c$ ($Y \leq: c$). Rule Getfield~(r1.b) handles the situation where $Y$ has an associated symbolic object in the heap, but this does not contain the field $f$. If we assume the programs to be type-correct, this situation may only arise if the symbolic object was previously created with the help of a superclass of $c$. Note that, upon creation, a symbolic object is assumed to possibly be of \emph{any subclass} of its creation class. If later during symbolic execution a finer type assumption must be performed, as in this case, the missing fields are added to the object, with all the new slots left uninitialized. Also, the path condition is updated to reflect the new type assumption on the symbolic object bound to $Y$ ($Y \leq: c$). Finally, the rule Getfield~(r2) handles the situation where the field $f$ in the symbolic object pointed by $Y$ is uninitialized ($H(Y)(f) = \bot$). The refinement rule therefore initializes the field with a special value $assume(H, Y, f, Z)$ that accounts for all the possible alias relations between the symbolic reference $Y$ and all the other symbolic references that are bound. Formally the function $assume(H, Y, f, Z)$ is defined as follows: 
\begin{multline*}    
assume(H, Y, f, Z) = \code{ite(} Y = Y_1, H(Y_1)(f), \ldots, \\
\code{ite(} Y = Y_m, H(Y_m)(f), Z \code{)} \ldots \code{)}
\end{multline*}
where $\{Y_1 \ldots Y_m\} = \{ Y' \in \mathrm{dom}(H) \ | \ Y' \neq Y \, \land \, f \in \mathrm{dom}(H(Y')) \, \land\, H(Y')(f) \neq \bot \}$. The set $\{Y_1 \ldots Y_m\}$ is the \emph{alias set} of $Y$, and is composed by the set of all the bound symbolic references that point to a type-compatible object with $c$, and whose $f$ field is initialized. The symbolic value $Z$ stands for the initial value of $Y.f$, the value that would have the field $Y.f$ if symbolic execution up to the current instant of time did not modify the field. The $assume$ value that is assigned to the $Y.f$ field reflects the possible current value of the field: For all $Y_i$, If $Y$ aliases $Y_i$, then $Y.f$ has the value stored in $Y_i.f$. If $Y$ does not alias any object in its alias set, then it still contains the unmodified initial value $Z$.

\begin{figure}[tp]
\centering
\includegraphics[width=0.45\textwidth]{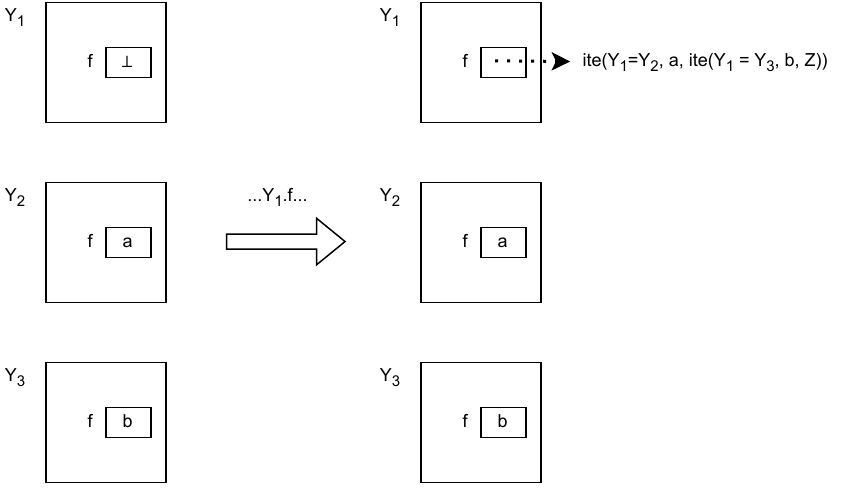}
\caption{Example semantics of field reading expressions}\label{fig:pose-getfield}
\end{figure}

Figure~\ref{fig:pose-getfield} reports an example illustrating the effect of the evaluation of a field reading expression. Let us suppose that the field $Y_1.f$ is uninitialized, that the expression $Y_1.f$ must be evaluated, and that two symbolic references $Y_2$ and $Y_3$ are bound in the heap with a value for field $f$. The effect of refinement is to initialize field $Y_1.f$ to the \code{ite} value shown in figure, which takes into account the possible alias relations of $Y_1$ with $Y_2$ and $Y_3$. This value is returned as a result of the evaluation of the expression.

The computation semantics of field assignment expressions, $e.f := e$, can be described as follows. Rule Putfield~(c5) is for concrete objects: If $l$ is the location of a concrete object, the field $f$ of the object is directly updated. More complex is the effect of updating a field of a symbolic object, as described by rule Putfield~(c1). The update can be performed only if $Y$ is bound and the symbolic object has field $f$: Otherwise, refinement is necessary: The corresponding refinement transitions, Putfield~(r1) and Putfield~(r1.b), are almost identical to Getfield~(r1) and Getfield~(r1.b), and will not be discussed. The effect of a field update does not only affect the symbolic object bound to $Y$: It affects all the symbolic objects in the alias set of Y. Formally the impact of the update on the alias set is described by the definition of $H_r = update(H, Y, f, \sigma')$: 
\begin{multline*}    
\mathrm{dom}(H_r)=\mathrm{dom}(H) \, \wedge \, \forall \,u' \in \mathrm{dom}(H), \, f' \in \mathrm{dom}(H)(Y') \,| \phantom{} \\ u' = Y' \, \land \, Y' \neq Y \, \land \, f' = f \, \land \, H(Y')(f) \neq \bot\\ \implies H_r(u')(f') = \code{ite(}Y = Y', \sigma', H(u')(f')\code{)} \, \wedge \\
\neg(u' = Y' \, \land \, Y' \neq Y \, \land \, f' = f \, \land \, H(Y')(f) \neq \bot)\\ \implies H_r(u')(f') = H(u')(f').
\end{multline*}    
In plain terms, if $Y'$ is a bound symbolic reference to a type-compatible object, whose $f$ field is initialized, the field $Y'.f$ must also be updated to an \code{ite} reflecting the potential alias relation: If $Y$ aliases $Y'$, then $Y'.f$ also assumes the value assigned to $Y.f$, otherwise it retains its current value. Finally, rules Putfield~(c2)-(c4) handle the case where an $\code{ite(}\sigma, \sigma_1, \sigma_2\code{)}$ symbolic reference is used to refer the object to be updated. Rules Putfield~(c3) and (c4) are analogous to Getfield~(c3) and (c4), and will not be discussed. Rule Putfield~(c2) defines the effects of a field update on the heap and the path condition in terms of a suitable merging between the effects of updating the objects referred by $\sigma_1$ and $\sigma_2$ respectively. The formal definitions of the $mergeHPf$ and $mergeClauses$ functions are as follows. It is $H' = mergeHPf(H'_1, H'_2, 
f, \sigma)$ iff  $\mathrm{dom}(H')=\mathrm{dom}(H'_1) \cup \mathrm{dom}(H'_2)$, and for all  $u \in \mathrm{dom}(H')$ it is:
\begin{enumerate}
\item if $u \in \mathrm{dom}(H'_1)$, and $u \notin \mathrm{dom}(H'_2)$, then $\mathrm{dom}(H'(u)) = \mathrm{dom}(H'_1(u))$, and for all $f \in \mathrm{dom}(H'(u))$:
\begin{enumerate}
\item if $H'_1(u)(f) = \bot$ then it is $H'(u)(f)= \bot$; 
\item otherwise it is $H'(u)(f)= \code{ite(} \sigma, \sigma_1, Z \code{)}$, where $\sigma_1 = H'_1(u)(f)$ and $Z$ is a fresh symbol;
\end{enumerate}
\item if $u \notin \mathrm{dom}(H'_1)$, and $u \in \mathrm{dom}(H'_2)$, then $\mathrm{dom}(H'(u)) = \mathrm{dom}(H'_2(u))$, and for all $f \in \mathrm{dom}(H'(u))$:
\begin{enumerate}
\item if $H'_2(u)(f) = \bot$ then it is $H'(u)(f)= \bot$; 
\item otherwise it is $H'(u)(f)=\code{ite(} \sigma, Z, \sigma_2 \code{)}$, where $\sigma_2 = H'_2(u)(f)$ and $Z$ is a fresh symbol;
\end{enumerate}
\item if $u \in \mathrm{dom}(H'_1)$, and $u \in \mathrm{dom}(H'_2)$, and $H'_1(u) = H'_2(u)$, then $H'(u) = H'_1(u) = H'_2(u)$;
\item if $u \in \mathrm{dom}(H'_1)$, and $u \in \mathrm{dom}(H'_2)$, and $H'_1(u) \neq H'_2(u)$, then $\mathrm{dom}(H'(u)) = \mathrm{dom}(H'_1(u)) \cup \mathrm{dom}(H'_2(u))$, and for all $f \in \mathrm{dom}(H'(u))$:
\begin{enumerate}
\item if $f \in \mathrm{dom}(H'_1(u))$ and $f \notin \mathrm{dom}(H'_2(u))$, then: 
\begin{enumerate}
\item if $H'_1(u)(f) = \bot$ then $H'(u)(f)=\bot$; 
\item otherwise, $H'(u)(f)=\code{ite(} \sigma, \sigma_1, Z \code{)}$, where $\sigma_1 = H'_1(u)(f)$ and $Z$ is a fresh symbol;
\end{enumerate}
\item if $f \in \mathrm{dom}(H'_2(u))$ and $f \notin \mathrm{dom}(H'_1(u))$, then: 
\begin{enumerate}
\item if $H'_2(u)(f) = \bot$ then $H'(u)(f)=\bot$; 
\item otherwise, $H'(u)(f)=\code{ite(} \sigma, Z, \sigma_2 \code{)}$, where $\sigma_2 = H'_2(u)(f)$ and $Z$ is a fresh symbol;
\end{enumerate}
\item if $f \in \mathrm{dom}(H'_1(u))$, and $f \in \mathrm{dom}(H'_2(u))$, and $H'_1(u)(f) = H'_2(u)(f)$ then $H'(u)(f)=H'_1(u)(f)=H'_2(u)(f)$;
\item if $f \in \mathrm{dom}(H'_1(u))$, and $f \in \mathrm{dom}(H'_2(u))$, and $H'_1(u)(f) \neq \bot$, and $H'_2(u)(f) = \bot$, then $H'(u)(f)=\code{ite(} \sigma, \sigma_1, Z \code{)}$, where $\sigma_1 = H'_1(u)(f)$ and $Z$ is a fresh symbol;
\item if $f \in \mathrm{dom}(H'_1(u))$, and $f \in \mathrm{dom}(H'_2(u))$, and $H'_1(u)(f) = \bot$, and $H'_2(u)(f) \neq \bot$, then $H'(u)(f)=\code{ite(} \sigma, Z, \sigma_2 \code{)}$, where $\sigma_2 = H'_2(u)(f)$ and $Z$ is a fresh symbol;
\item if $f \in \mathrm{dom}(H'_1(u))$, and $f \in \mathrm{dom}(H'_2(u))$, and $H'_1(u)(f) \neq H'_2(u)(f)$, and $H'_1(u)(f) \neq \bot$, and $H'_2(u)(f) \neq \bot$ then $H'(u)(f)=\code{ite(} \sigma, \sigma_1, \sigma_2 \code{)}$, where $\sigma_1 = H'_1(u)(f)$ and $\sigma_2 = H'_2(u)(f)$.
\end{enumerate}
\end{enumerate}
The function $mergeClauses(H'_1, H'_2, f, \sigma)$ yields the set of all the clauses $Y.f = Z$, where $Z$ are all (and only) the fresh symbols introduced in the definition of $mergeHPf$, and $Y$ is the symbolic input reference that points to the symbolic object where $Z$ has been injected.

\begin{figure}[tp]
\centering
\includegraphics[width=0.45\textwidth]{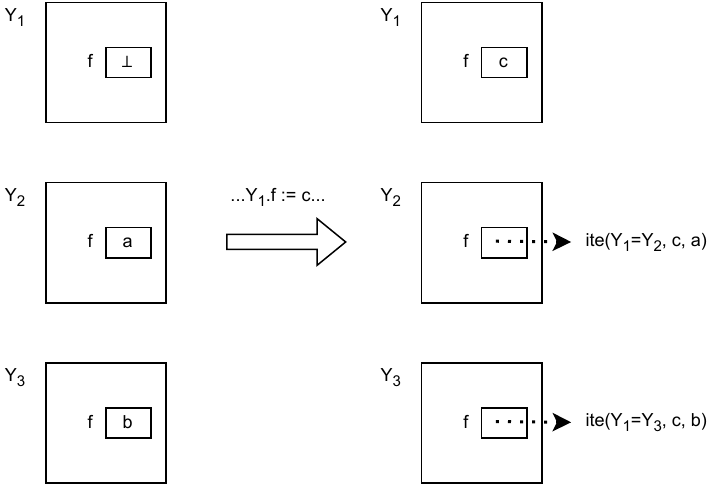}
\caption{Example semantics of field assignment expressions}\label{fig:pose-putfield}
\end{figure}

Figure~\ref{fig:pose-putfield} reports an example of the effect of the evaluation of a field assignment expression. We again suppose that three symbolic references, $Y_1$, $Y_2$, $Y_3$, are bound to symbolic objects, that all the symbolic objects define the field $f$, that the object $Y_2$ and $Y_3$ have field $f$ initialized, and that the expression $Y_1.f := c$ must be evaluated. In this situation no refinement transition fires: The computation transition updates the field $f$ of $Y_1$ to the $c$ value, and as a further effect updates the fields $f$ of all the other symbolic objects with \code{ite} terms that reflect the possible effect of the update on these objects in case an alias relation subsists.

Conditional expressions $\code{if\ } e\ e\ e$ have associated only computation transitions. Rules If~(c1) and If~(c2) manage the cases $\code{if\ true} \ldots$ and $\code{if\ false} \ldots$, that reduce to the evaluation of the corresponding branch. Rules If~(c3) and If~(c4) manage the case where the condition of the $\code{if}$ expression is symbolic. In this case both transitions may fire, and the path condition is updated to reflect the assumption associated to the branch taken.

Finally, method invocation expressions $e.m(\rep{e})$ are handled only by computation transitions. We introduce the following definitions:
\begin{itemize}
\item The function $recv(c, m)$ yields the class that provides the implementation of method $m$ to class $c$ (it is $c$ itself iff $c$ implements $m$, or an ancestor of $c$ iff $c$ inherits the implementation of $m$).
\item The predicate $impl(c, m)$ is true iff $c$ implements $m$. 
\item The predicate $sees(c, m, c')$ is true iff $c$ is a subclass of $c'$, $c'$ implements the method $m$, and no superclass of $c$ and subclass of $c'$ implements $m$ (note that $c$ may implement $m$). 
\item The predicate $overrides(c, m, c')$ is defined as $c \neq c' \land sees(c, m, c') \land impl(c, m)$. 
\end{itemize}
We can deduce that $recv(c, m) = c'$, for $c \neq c'$, iff $sees(c, m, c') \land \neg impl(c, m)$, and that $recv(c, m) = c$ iff $impl(c, m)$. Given the above definitions it is easy to see that rule Invoke~(c1) (method invocation on concrete objects) finds the implementation of the method of the object pointed by location $l$, and invokes it. Different is the case (rule Invoke~(c2)) where the receiver of the method invocation is a symbolic reference. In this case, since the symbolic reference might refer to an object of any class that declares the method $m$, symbolic execution branches to all possible implementations of $m$. The path condition at each branch is assumed to reflect the assumption on the class of the symbolic receiver object: When the method $m$ is executed, if $m$ is declared by class $c'$, the object may have any subclass of $c'$ that does not override $m$. The final rules Invoke~(c3) and Invoke~(c4) handle the case where the reference to the receiver object is an $\code{ite(} \sigma, \sigma_1, \sigma_2 \code{)}$ symbolic reference. In synthesis, when such a reference is encountered, symbolic execution branches.

We conclude this Appendix by briefly discussing the remaining rules. The Let computation rule has the effect of performing a term substitution in its main argument. The Op~(c1) and Op~(c2) computation rules yield the result of the application of a binary or unary operator to concrete operands. The Eq~(c1)-(c7) computation rules account for the semantics of equality, where syntactically identical values are always considered equal, distinct concrete locations are always considered different each other and from \code{null}, and symbolic references are always considered different from concrete locations. Finally, the context rules Ctx~(c) and Ctx~(r) extend the reduction semantics of expressions under the admissible evaluation contexts.

%% file: bibliography.bib
@inproceedings{yang2012memoized,
  title={Memoized symbolic execution},
  author={Yang, Guowei and P{\u{a}}s{\u{a}}reanu, Corina S and Khurshid, Sarfraz},
  booktitle={Proceedings of the 2012 International Symposium on Software Testing and Analysis},
  pages={144--154},
  year={2012}
}

@article{yang:incremental:tosem:2014,
author = {Yang, Guowei and Person, Suzette and Rungta, Neha and Khurshid, Sarfraz},
title = {Directed Incremental Symbolic Execution},
year = {2014},
issue_date = {September 2014},
publisher = {Association for Computing Machinery},
address = {New York, NY, USA},
volume = {24},
number = {1},
issn = {1049-331X},
doi = {10.1145/2629536},
journal = {ACM Trans. Softw. Eng. Methodol.},
month = {oct},
articleno = {3},
numpages = {42}
}

@INPROCEEDINGS{filieri:probabilistic:ase:2015,
  author={Filieri, Antonio and Pasareanu, Corina S. and Yang, Guowei},
  booktitle={2015 30th IEEE/ACM International Conference on Automated Software Engineering (ASE)}, 
  title={Quantification of Software Changes through Probabilistic Symbolic Execution (N)}, 
  year={2015},
  volume={},
  number={},
  pages={703-708},
  doi={10.1109/ASE.2015.78}
}

@ARTICLE{he:se:smartcontract:2020,
  author={He, Daojing and Deng, Zhi and Zhang, Yuxing and Chan, Sammy and Cheng, Yao and Guizani, Nadra},
  journal={IEEE Network}, 
  title={Smart Contract Vulnerability Analysis and Security Audit}, 
  year={2020},
  volume={34},
  number={5},
  pages={276-282},
  doi={10.1109/MNET.001.1900656}}

@inproceedings{zheng2022park,
  title={Park: Accelerating smart contract vulnerability detection via parallel-fork symbolic execution},
  author={Zheng, Peilin and Zheng, Zibin and Luo, Xiapu},
  booktitle={Proceedings of the 31st ACM SIGSOFT International Symposium on Software Testing and Analysis},
  pages={740--751},
  year={2022}
}

@article{symbex:fuzzing:2011,
author = {Wang, Tielei and Wei, Tao and Gu, Guofei and Zou, Wei},
title = {Checksum-Aware Fuzzing Combined with Dynamic Taint Analysis and Symbolic Execution},
year = {2011},
issue_date = {September 2011},
publisher = {Association for Computing Machinery},
address = {New York, NY, USA},
volume = {14},
number = {2},
issn = {1094-9224},
doi = {10.1145/2019599.2019600},
journal = {ACM Trans. Inf. Syst. Secur.},
month = {sep},
articleno = {15},
numpages = {28}
}

@inproceedings{stephens2016driller,
  title={Driller: Augmenting fuzzing through selective symbolic execution.},
  author={Stephens, Nick and Grosen, John and Salls, Christopher and Dutcher, Andrew and Wang, Ruoyu and Corbetta, Jacopo and Shoshitaishvili, Yan and Kruegel, Christopher and Vigna, Giovanni},
  booktitle={NDSS},
  volume={16},
  pages={1--16},
  year={2016}
}

@inproceedings{pasareanu:mixed:issta:2011,
  author    = {Corina S. P\v{a}s\v{a}reanu and
               Neha Rungta and
               Willem Visser},
  title     = {Symbolic execution 
with mixed concrete-symbolic solving},
  booktitle = {International Symposium on Software Testing and Analysis},
  year      = {2011},
  pages_hidden     = {34-44},
  ee_hidden        = {http://doi.acm.org/10.1145/2001420.2001425},
  }

@inproceedings{sinha:symbolic:fmcad:2008,
 author = {Nishant Sinha},
 title = {Symbolic program analysis using term rewriting and generalization},
 booktitle = {International Conference on Formal Methods in Computer-Aided Design},
 series_hidden = {FMCAD '08},
 year = {2008},
 location_hidden = {Portland, Oregon},
 pages = {19:1--19:9},
 publisher = {IEEE},
}

@Article{kiasan:jase:2012,
  author = 	 {Deng, Xianghua and Lee, Jooyong and Robby},
  title = 	 {Efficient and formal generalized symbolic execution},
  journal = 	 {Automated Software Engineering},
  year = 	 {2012},
  doi = {10.1007/s10515-011-0089-9},
  volume = 	 {19},
  pages = 	 {233--301},
}

@INPROCEEDINGS{laziersharp:sefm:2007,
  author={Deng, Xianghua and Robby and Hatcliff, John},
  booktitle={Fifth IEEE International Conference on Software Engineering and Formal Methods (SEFM 2007)}, 
  title={Towards A Case-Optimal Symbolic Execution Algorithm for Analyzing Strong Properties of Object-Oriented Programs}, 
  year={2007},
  volume={},
  number={},
  pages={273-282},
  doi={10.1109/SEFM.2007.43}
}

@inproceedings{symbtypes:cc:2017,
author = {Li, Lian and Lu, Yi and Xue, Jingling},
title = {Dynamic Symbolic Execution for Polymorphism},
year = {2017},
isbn = {9781450352338},
publisher = {Association for Computing Machinery},
address = {New York, NY, USA},
doi = {10.1145/3033019.3033029},
booktitle = {Proceedings of the 26th International Conference on Compiler Construction},
pages = {120–130},
numpages = {11},
location = {Austin, TX, USA},
series = {CC 2017}
}

@InProceedings{khurshid:tacas:2003,
  author =   {S. Khurshid and C. S. P\u{a}s\u{a}reanu and
                  W. Visser},
  title =    {Generalized Symbolic Execution for Model Checking
                  and Testing},
  booktitle =    {Tools and Algorithms
                  for Construction and Analysis of Systems},
  pages_hidden =    {553--568},
  year =     2003,
  publisher =   {Springer},
  series =   {LNCS 2619}
}

@inproceedings{braione:enhancing:esecfse:2013,
 author = {Braione, Pietro and Denaro, Giovanni and Pezz\`{e}, Mauro},
 title = {Enhancing Symbolic Execution with Built-in Term Rewriting and Constrained Lazy Initialization},
 booktitle = {Proc. of the 2013 9th Joint Meeting on Foundations of Software Engineering},
 series = {ESEC/FSE 2013},
 year = {2013},
 isbn = {978-1-4503-2237-9},
 location = {Saint Petersburg, Russia},
 pages = {411--421},
 numpages = {11},
 doi = {10.1145/2491411.2491433},
 acmid = {2491433},
 publisher = {ACM},
 address = {New York, NY, USA},
}

@article{braione_software_2014,
  title = {Software {{Testing}} with {{Code}}-{{Based Test Generators}}: {{Data}} and {{Lessons Learned}} from a {{Case Study}} with an {{Industrial Software Component}}},
  volume = {22},
  number = {2},
  journal = {Software Quality Journal},
  author = {Braione, Pietro and Denaro, Giovanni and Mattavelli, Andrea and Vivanti, Mattia and Muhammad, Ali},
  month = jun,
  year = {2014},
  pages = {311-333}
}

@article{king_symbolic_1976,
  title = {Symbolic {{Execution}} and {{Program Testing}}},
  volume = {19},
  number = {7},
  journal = {Communications of the ACM},
  author = {King, James C.},
  year = {1976},
  pages = {385-394}
}

@inproceedings{deng_bogor_kiasan_2006,
  title = {Bogor/{{Kiasan}}: {{A K}}-{{Bounded Symbolic Execution}} for {{Checking Strong Heap Properties}} of {{Open Systems}}},
  booktitle = {{{IEEE}}/{{ACM International Conference}} on {{Automated Software Engineering}} ({{ASE}})},
  author = {Deng, Xianghua and Lee, Jooyong and {Robby}},
  year = {2006},
  pages = {157-166}
}

@inproceedings{clarke_program_1976,
  title = {A {{Program Testing System}}},
  booktitle = {{{ACM Annual Conference}}},
  author = {Clarke, Lori A.},
  year = {1976},
  pages = {488-491}
}

@inproceedings{burnim_wise_2009,
  title = {{{WISE}}: {{Automated Test Generation}} for {{Worst}}-{{Case Complexity}}},
  booktitle = {{{IEEE}}/{{ACM International Conference}} on {{Software Engineering}} ({{ICSE}})},
  author = {Burnim, Jacob and Juvekar, Sudeep and Sen, Koushik},
  year = {2009},
  pages = {463--473}
}

@article{cadar_symbolic_2013,
  title = {Symbolic {{Execution}} for {{Software Testing}}: {{Three Decades Later}}},
  volume = {56},
  number = {2},
  journal = {Communications of the ACM},
  author = {Cadar, Cristian and Sen, Koushik},
  month = feb,
  year = {2013},
  pages = {82-90}
}

@inproceedings{luckow_symbolic_2017,
  title = {Symbolic {{Complexity Analysis Using Context}}-{{Preserving Histories}}},
  booktitle = {{{IEEE International Conference}} on {{Software Testing}}, {{Verification}} and {{Validation}} ({{ICST}})},
  author = {Luckow, Kasper and Kersten, Rody and P\u{a}s\u{a}reanu, Corina},
  month = mar,
  year = {2017},
  pages = {58-68}
}

@inproceedings{tillmann_pex_2008,
  title = {Pex: {{White Box Test Generation}} for .{{NET}}},
  booktitle = {International {{Conference}} on {{Tests}} and {{Proofs}} ({{TAP}})},
  author = {Tillmann, Nikolai and {de Halleux}, Jonathan},
  year = {2008},
  pages = {134-153}
}

@inproceedings{anand_jpf-se_2007,
  title = {{{JPF}}-{{SE}}: {{A Symbolic Execution Extension}} to {{Java PathFinder}}},
  booktitle = {International {{Conference}} on {{Tools}} and {{Algorithms}} for {{Construction}} and {{Analysis}} of {{Systems}} ({{TACAS}})},
  author = {Anand, Saswat and P\u{a}s\u{a}reanu, Corina S. and Visser, Willem},
  year = {2007},
  pages = {134-138}
}

@inproceedings{de_moura_z3_2008,
  title = {Z3: {{An Efficient SMT Solver}}},
  booktitle = {International {{Conference}} on {{Tools}} and {{Algorithms}} for {{Construction}} and {{Analysis}} of {{Systems}} ({{TACAS}})},
  author = {De Moura, Leonardo and Bj{\o}rner, Nikolaj},
  year = {2008},
  pages = {337-340}
}

@inproceedings{braione_symbolic_2015,
  title = {Symbolic {{Execution}} of {{Programs}} with {{Heap Inputs}}},
  booktitle = {{{ACM Joint European Software Engineering Conference}} and {{Symposium}} on the {{Foundations}} of {{Software Engineering}} ({{ESEC}}/{{FSE}})},
  author = {Braione, Pietro and Denaro, Giovanni and Pezz{\`e}, Mauro},
  year = {2015},
  pages = {602-613}
}

@inproceedings{vivanti_search-based_2013,
  title = {Search-{{Based Data}}-{{Flow Test Generation}}},
  booktitle = {{{IEEE International Symposium}} on {{Software Reliability Engineering}} ({{ISSRE}})},
  author = {Vivanti, Mattia and Mis, Andre and Gorla, Alessandra and Fraser, Gordon},
  year = {2013},
  pages = {370-379}
}

@inproceedings{li_klover_2011,
  title = {{{KLOVER}}: {{A Symbolic Execution}} and {{Automatic Test Generation Tool}} for {{C}}++ {{Programs}}},
  booktitle = {International {{Conference}} on {{Computer Aided Verification}} ({{CAV}})},
  author = {Li, Guodong and Ghosh, Indradeep and Rajan, Sreeranga},
  year = {2011},
  pages = {53-68}
}

@inproceedings{cadar_klee_2008,
  title = {{{KLEE}}: {{Unassisted}} and {{Automatic Generation}} of {{High}}-Coverage {{Tests}} for {{Complex Systems Programs}}},
  booktitle = {Symposium on {{Operating Systems Design}} and {{Implementation}} ({{OSDI}})},
  author = {Cadar, Cristian and Dunbar, Daniel and Engler, Dawson},
  year = {2008},
  pages = {209-224}
}

@inproceedings{braione_jbse_2016,
  title = {{{JBSE}}: {{A Symbolic Executor}} for {{Java Programs}} with {{Complex Heap Inputs}}},
  booktitle = {{{ACM Joint European Software Engineering Conference}} and {{Symposium}} on the {{Foundations}} of {{Software Engineering}} ({{ESEC}}/{{FSE}})},
  author = {Braione, Pietro and Denaro, Giovanni and Pezz{\`e}, Mauro},
  year = {2016},
  pages = {1018-1022}
}

@inproceedings{burnim_heuristics_2008,
  title = {Heuristics for {{Scalable Dynamic Test Generation}}},
  booktitle = {{{IEEE}}/{{ACM International Conference}} on {{Automated Software Engineering}} ({{ASE}})},
  author = {Burnim, Jacob and Sen, Koushik},
  year = {2008},
  pages = {443-446}
}

@inproceedings{geldenhuys2013bounded,
  title={Bounded lazy initialization},
  author={Geldenhuys, Jaco and Aguirre, Nazareno and Frias, Marcelo F and Visser, Willem},
  booktitle={NASA Formal Methods: 5th International Symposium, NFM 2013, Moffett Field, CA, USA, May 14-16, 2013. Proceedings 5},
  pages={229--243},
  year={2013},
  organization={Springer}
}

@article{rosner2015bliss,
  title={BLISS: improved symbolic execution by bounded lazy initialization with SAT support},
  author={Rosner, Nicol{\'a}s and Geldenhuys, Jaco and Aguirre, Nazareno M and Visser, Willem and Frias, Marcelo F},
  journal={IEEE Transactions on Software Engineering},
  volume={41},
  number={7},
  pages={639--660},
  year={2015},
  publisher={IEEE}
}

@article{baluda2016bidirectional,
  title={Bidirectional symbolic analysis for effective branch testing},
  author={Baluda, Mauro and Denaro, Giovanni and Pezz{\`e}, Mauro},
  journal={IEEE Transactions on Software Engineering},
  volume={42},
  number={5},
  pages={403--426},
  year={2016},
  publisher={IEEE}
}

@inproceedings{aquino_worst-case_2018,
  title = {Worst-{{Case Execution Time Testing}} via {{Evolutionary Symbolic Execution}}},
  copyright = {All rights reserved},
  booktitle = {International {{Symposium}} on {{Software Reliability Engineering}} ({{ISSRE}})},
  author = {Aquino, Andrea and Denaro, Giovanni and Salza, Pasquale},
  month = oct,
  year = {2018},
  pages = {76-87}
}

@inproceedings{visser:test:issta:2004,
	Author = {Willem Visser and Corina S.~P{\u a}s{\u a}reanu and Sarfraz Khurshid},
	Bibsource = {DBLP, http://dblp.uni-trier.de},
	Booktitle = {Proceedings of the 2004 ACM SIGSOFT International Symposium on Software Testing and Analysis (ISSTA 2004)},
	Pages = {97--107},
	Publisher = {ACM},
	Title = {Test input generation with {Java PathFinder}},
	Year = {2004}}

@inproceedings{bucur:parallel:eurosys:2011,
	Author = {Stefan Bucur and Vlad Ureche and Cristian Zamfir and George Candea},
	Booktitle = {Proceedings of EuroSys 2011},
	Publisher = {ACM},
	Title = {Parallel Symbolic Execution for Automated Real-World Software Testing},
	Year = 2011}

@inproceedings{godefroid:dart:pldi:2005,
	Author = {Patrice Godefroid and Nils Klarlund and Koushik Sen},
	Booktitle = {Proceedings of the ACM SIGPLAN 2005 Conference on Programming Language Design and Implementation (PLDI 2005)},
	Pages = {213--223},
	Title = {{DART}: directed automated random testing},
	Year = {2005}}

@inproceedings{sen:cute:esec:2005,
	Author = {Koushik Sen and Darko Marinov and Gul Agha},
	Booktitle = {Proceedings of the 10th European software engineering conference held jointly with 13th ACM SIGSOFT international symposium on Foundations of software engineering (ESEC/FSE-13)},
	Pages = {263--272},
	Title = {{CUTE}: a concolic unit testing engine for {C}},
	Year = 2005}

@inproceedings{coen:symbolic:fse:2001,
	Author = {A. Coen-Porisini and G. Denaro and C. Ghezzi and M. Pezz{\`e}},
	Booktitle = {Proceedings of the Joint European Software Engineering Conference and ACM SIGSOFT Symposium on the Foundations of Software Engineering (ESEC/FSE 2001)},
	Pages = {142--151},
	Title = {Using symbolic execution for verifying safety-critical systems},
	Year = {2001}}

@inproceedings{staats:parallel:issta:2010,
	Author = {Staats, Matt and P\u{a}s\u{a}reanu, Corina S.},
	Booktitle = {Proceedings of the 19th International Symposium on Software Testing and Analysis (ISSTA 2010)},
	Doi = {10.1145/1831708.1831732},
	Isbn = {978-1-60558-823-0},
	Pages = {183--194},
	Publisher = {ACM},
	Title = {Parallel Symbolic Execution for Structural Test Generation},
	Year = {2010}}

@inproceedings{siddiqui:staged:sac:2012,
 author = {Siddiqui, Junaid Haroon and Khurshid, Sarfraz},
 title = {Staged symbolic execution},
 booktitle = {ACM Symposium on Applied Computing},
 series_hidden = {SAC '12},
 year = {2012},
 isbn = {978-1-4503-0857-1},
 location_hidden = {Trento, Italy},
 pages_hidden = {1339--1346},
 numpages = {8},
 url_hidden = {http://doi.acm.org/10.1145/2231936.2231988},
 doi = {10.1145/2231936.2231988},
 acmid = {2231988},
 publisher_hidden = {ACM},
 address_hidden = {New York, NY, USA},
}

@inproceedings{statemerging:pldi:2012,
author = {Kuznetsov, Volodymyr and Kinder, Johannes and Bucur, Stefan and Candea, George},
title = {Efficient State Merging in Symbolic Execution},
year = {2012},
isbn = {9781450312059},
publisher = {Association for Computing Machinery},
address = {New York, NY, USA},
doi = {10.1145/2254064.2254088},
booktitle = {Proceedings of the 33rd ACM SIGPLAN Conference on Programming Language Design and Implementation},
pages = {193–204},
numpages = {12},
location = {Beijing, China},
series = {PLDI '12}
}

@InProceedings{statejoining:rv:2009,
author="Hansen, Trevor
and Schachte, Peter
and S{\o}ndergaard, Harald",
editor="Bensalem, Saddek
and Peled, Doron A.",
title="State Joining and Splitting for the Symbolic Execution of Binaries",
booktitle="Runtime Verification",
year="2009",
publisher="Springer Berlin Heidelberg",
address="Berlin, Heidelberg",
pages="76--92",
isbn="978-3-642-04694-0"
}

@inproceedings{efficient:date:2008,
author = {Arons, Tamarah and Elster, Elad and Ozer, Shlomit and Shalev, Jonathan and Singerman, Eli},
title = {Efficient Symbolic Simulation of Low Level Software},
year = {2008},
isbn = {9783981080131},
publisher = {Association for Computing Machinery},
address = {New York, NY, USA},
doi = {10.1145/1403375.1403577},
booktitle = {Proceedings of the Conference on Design, Automation and Test in Europe},
pages = {825–830},
numpages = {6},
location = {Munich, Germany},
series = {DATE '08}
}

@phdthesis{babic:phd:2008,
  author = {Babi\'c, Domagoj},
  title = {{Exploiting Structure for Scalable Software Verification}},
  school = {University of British Columbia, Vancouver, Canada},
  year = {2008},
}

@article{constructing:ijpp:2005,
author = {Koelbl, Alfred and Pixley, Carl},
title = {Constructing Efficient Formal Models from High-Level Descriptions Using Symbolic Simulation},
 journal = {International Journal of Parallel Programming},
 volume = {33},
 number = {6},
 month = dec,
 year = {2005},
 issn = {1573-7640},
 pages = {645--666},
 doi = {10.1007/s10766-005-8910-3},
}

@article{SurveySymExec-CSUR18,
  author    = {Baldoni, Roberto and Coppa, Emilio and D'Elia, Daniele Cono and Demetrescu, Camil and Finocchi, Irene},
  title     = {A Survey of Symbolic Execution Techniques},
  journal   = {ACM Comput. Surv.},
  volume    = {51},
  number = {3},
  articleno = {50},
  publisher = {ACM},
  address = {New York, NY, USA},
  year = {2018}
}

@article{Pasareanu:survey:JSTTT2009,
 author = {Pasareanu, Corina S. and Visser, Willem},
 title = {A Survey of New Trends in Symbolic Execution for Software Testing and Analysis},
 journal = {International Journal on Software Tools for Technology Transf.},
 issue_date = {October 2009},
 volume = {11},
 number = {4},
 month = oct,
 year = {2009},
 issn = {1433-2779},
 pages = {339--353},
 numpages = {15},
 doi = {10.1007/s10009-009-0118-1},
 acmid = {1667869},
 publisher = {Springer-Verlag},
 address = {Berlin, Heidelberg},
}

@INPROCEEDINGS{panichella:comp:sbst:2017,
  author={Panichella, Annibale and Molina, Urko Rueda},
  booktitle={2017 IEEE/ACM 10th International Workshop on Search-Based Software Testing (SBST)}, 
  title={Java Unit Testing Tool Competition - Fifth Round}, 
  year={2017},
  pages={32--38},
  doi={10.1109/SBST.2017.7}
}

@inproceedings{sen:multise:fse:2015,
author = {Sen, Koushik and Necula, George and Gong, Liang and Choi, Wontae},
title = {MultiSE: multi-path symbolic execution using value summaries},
year = {2015},
isbn = {9781450336758},
publisher = {Association for Computing Machinery},
address = {New York, NY, USA},
url = {https://doi.org/10.1145/2786805.2786830},
doi = {10.1145/2786805.2786830},
booktitle = {Proceedings of the 2015 10th Joint Meeting on Foundations of Software Engineering},
pages = {842–853},
numpages = {12},
location = {Bergamo, Italy},
series = {ESEC/FSE 2015}
}

@inproceedings{hillery:summaries:vmcai:2016,
author = {Hillery, Benjamin and Mercer, Eric and Rungta, Neha and Person, Suzette},
title = {Exact Heap Summaries for Symbolic Execution},
year = {2016},
isbn = {9783662491218},
publisher = {Springer-Verlag},
address = {Berlin, Heidelberg},
url = {https://doi.org/10.1007/978-3-662-49122-5_10},
doi = {10.1007/978-3-662-49122-5\_10},
booktitle = {Proceedings of the 17th International Conference on Verification, Model Checking, and Abstract Interpretation},
pages = {206–225},
numpages = {20},
location = {St. Petersburg, FL, USA},
series = {VMCAI 2016}
}
